\documentclass[a4paper, 11pt]{article}
\usepackage{amssymb,amsmath}
\usepackage[T1]{fontenc} % if needed
\usepackage[utf8]{inputenc}
\usepackage[english]{babel}
\usepackage{array}
\usepackage{booktabs} 	% hdashline
\usepackage{caption}
\usepackage{graphicx}
\usepackage{subfig}	% sottocaption
\usepackage{authblk}
\usepackage{arydshln} 	% hdashline
\usepackage{mathrsfs}	% mathscr per F (1.15)
\usepackage{multirow}	% tabella csec OS
\usepackage{siunitx}
\usepackage{graphicx,rotating,booktabs}
\usepackage{comment}

\usepackage[top=1.25in, bottom=1.75in,
left=1.25in, right=1.25in]{geometry}

\newcommand{\Pnue}{\nu_{\mbox{\tiny e}}}
\newcommand{\APnue}{\overline{\nu}_{\!\;\!\mbox{\tiny e}}}
\newcommand{\Pnux}{\nu_{\mbox{\tiny $x$}}}

\newcommand\Tstrut{\rule{0pt}{2.6ex}}  % = `top' strut
\newcommand\Bstrut{\rule[-1.2ex]{0pt}{0pt}}% = `bottom'
\newcolumntype{R}{>{$}r<{$}}
\newcolumntype{L}{>{$}l<{$}}
\newcolumntype{C}{>{$}c<{$}}
\newcolumntype{M}{R@{${}\in{}$}L}
\newcolumntype{A}{R@{${}\pm{}$}L}
\DeclareSIUnit\erg{erg}
\DeclareSIUnit\parsec{pc}
\DeclareSIUnit\ton{ton}
\usepackage{hyperref}
\hypersetup{
    bookmarks=true,         % show bookmarks bar?
    unicode=false,          % non-Latin char in Acrobat’s bookmarks
    pdftoolbar=true,        % show Acrobat’s toolbar?
    pdfmenubar=true,        % show Acrobat’s menu?
    pdffitwindow=false,     % window fit to page when opened
    pdfstartview={FitH},    % fits the width of the page to the window
    pdfauthor={Kevin Multani},     % author
    colorlinks=true,       % false: boxed links; true: colored links
    linkcolor=blue,          % color of internal links
    citecolor=red,        % color of links to bibliography
}

\title{What can we learn on supernova neutrino spectra\\
with water Cherenkov detectors?}
\author[1,2,3]{A.\ Gallo Rosso\thanks{Corresponding
author: \href{mailto:andrea.gallorosso@gssi.it}
{andrea.gallorosso@gssi.it}.}}
\author[1,2]{F.\ Vissani}
\author[3]{M.C.\ Volpe}
\affil[1]{Gran Sasso Science Institute,
Viale F.\ Crispi 7, L'Aquila, Italy}
\affil[2]{INFN,
Laboratori Nazionali del Gran Sasso,%\\\hspace{\textwidth}
Via G. Acitelli, 22, Assergi, L'Aquila, Italy}
\affil[3]{Astro-Particule
et Cosmologie (APC),
CNRS UMR 7164, Universit\'e Denis Diderot\\\hspace{\textwidth}
10, rue Alice Domon et L\'eonie Duquet,
75205 Paris Cedex 13, France}
\date{}                     %% if you don't need date to appear
\begin{document}
	\maketitle
  \begin{abstract}
  We investigate the precision with which the supernova
neutrino spectra can be reconstructed in water Cherenkov detectors,
in particular the large scale Hyper-Kamiokande and Super-Kamiokande.
To this aim, we consider quasi-thermal neutrino spectra modified
by the Mikheev-Smirnov-Wolfenstein effect for the case of normal
ordering.
We perform three 9 degrees of freedom likelihood analyses
including first inverse-beta decay only, then the combination of
inverse beta decay and elastic scattering on electrons and finally
a third analysis that also includes neutral scattering
neutrino-oxygen events. A tenth parameter is added in the
analyses to account for the theoretical
uncertainty on the neutral current neutrino-oxygen cross section.
By assuming a 100\% efficiency in Hyper-Kamiokande,
we show that one can
reconstruct the electron antineutrino average energy and pinching
parameter with an accuracy of $\sim2\%$ and $\sim7\%$ percent
respectively, while
the antineutrino integrated luminosity can be pinned down at $\sim3\%$
percent level. As for the muon and tau neutrinos, the average
energy and the integrated luminosity can be measured
with $\sim7\%$ precision. 
These results represent a significant improvement with respect
Super-Kamiokande, particularly for the pinching parameter defining
the electron antineutrino spectra. 
As for electron neutrinos, the determination of the
emission parameters
requires the addition of supplementary detection channels.	
  \end{abstract}

\section{Introduction}
Unravelling the mechanisms for iron core-collapse supernovae is a
longstanding open question in astrophysics. 
The main ideas on the explosion mechanism were already outlined
in the sixties when
it was first  accepted that a massive star would undergo a
core collapse at the end of its life. According to Colgate
and Johnson's
prompt-shock model, core bounce would create a shock that would
propagate and expel the star's mantle \cite{Colgate:1960zz}.
Colgate and White suggested that the gravitational binding energy of
the newly formed proto-neutron star would be released mostly
as neutrinos, in a short burst lasting few seconds. A fraction of 
the order of a few percent of this energy could 
be deposited  behind the shock and drive the explosion
\cite{Colgate:1966ax}. This turned into the delayed-accretion shock
paradigm, proposed in the work of Wilson \cite{Wilson:1971}
and discussed since then, beginning with the works of Nadyozhin
\cite{dima} and of Bethe and Wilson \cite{Bethe:1984ux}.

Core-collapse supernova neutrinos have been first observed during
SN 1987A in the Large Magellanic Cloud through inverse beta decay in
the Kamiokande  \cite{Hirata:1987hu}, IMB \cite{Bionta:1987qt}
and Baksan detectors \cite{Alekseev:1988gp}. The LSD/Mont Blanc
data occurring a few hours before the others 
\cite{Dadykin:1987ek} have been debated (see e.g.\ \cite{jb}).
The time dependence of twenty-nine neutrino events 
has supported the delayed-accretion shock model, in contrast to
the prompt one, as shown by the in-depth 
analysis of Loredo and Lamb \cite{Loredo:2001rx} and
as confirmed by the completely 
independent analysis of ref.\ \cite{Pagliaroli:2008ur}. 
Based on the hypothesis of energy equipartition among the
neutrino flavors, the gravitational binding energy of the newly
formed neutron star is found to be approximatively  
\SI{3e53}{\erg},
in agreement with  predictions. Moreover, 
the SN 1987A neutrino spectra  do not contradict the hypothesis  
of quasi-thermal distribution, with emission temperatures of about
\SI{4}{\mega\electronvolt}   
(see \cite{Vissani:2014doa} for a systematic investigation). 

State-of-the-art supernova simulations are multi-dimensional,
include realistic neutrino transport, nuclear networks, convection
and hydrodynamic instabilities such as the
Standing-Accretion-Shock-Instability (SASI) \cite{Foglizzo:2006fu}.
The detection of the supernova neutrino time signal from a future
explosion will represent a direct  test of the SASI-aided
delayed-accretion shock mechanism (see e.g.\ \cite{Mueller:2012is}).
This is thought to explain the majority of supernova explosions,
except the most violent ones that might be magneto-hydrodynamically
driven \cite{Janka:2017vcp}. Successful explosions have been
obtained in two\babelhyphen{-}dimensions for a set of progenitors.
Three-dimensional simulations appear to be at the verge of succeed
(see e.g.\ \cite{Janka:2017vcp,Hix:2016qoa}). 

The occurrence of a supernova explosion in the Milky Way will 
trigger a network of detectors, based on water (or ice) Cherenkov,
scintillators and liquid argon detectors, providing the
collection of several thousands up to one million events for a
prototypical event located at $5\div\SI{10}{\kilo\parsec}$
from the Earth. The precise
knowledge of cross sections associated with inverse beta decay
and neutrino-electron scattering will allow  measurements, free
from the systematic errors that still affect neutrino-nucleus
cross sections in the several tens of MeV energy range. Neutrino
interaction cross sections on nuclei like oxygen, argon,
iron and lead at the relevant 
supernova energies have significant theoretical uncertainties.
Measurements are being planned
at the SNS facility by the COHERENT Collaboration \cite{Scholberg}.
These will bring important information on the reaction cross
sections and hopefully also on the quenching of the axial-vector
coupling constant, that is essential for neutrinoless
double beta decay.

According to supernova simulations, the neutrino spectra
in the decoupling region
are described by quasi-thermal
distributions. Three parameters for both electron and
non-electron type neutrinos are necessary to describe the spectra,
namely the normalization, the average energy and the width. 
The latter parameter describes the deviation from a perfect thermal
distribution, and it is commonly called  the ``pinching'' parameter.
This amounts to a set of 9 free parameters.\footnote{
We neglect possible spectral differences between muon and tau
(anti)neutrinos due to radiative corrections or new
non-standard interactions.} The situation gets more complex when
neutrinos enter the free-streaming regime, since the occurrence
of flavor conversion phenomena due to neutrino self-interactions,
coupling to matter, shock wave effects or turbulence might
produce spectral swapping(s).  In particular, neutrino
self-interactions make neutrino propagation in dense environments
a non-linear many-body problem. It still needs to be determined
under which conditions neutrino-neutrino
interactions produce spectral modifications,
and how sizable they are \cite{Volpe:2016bkp}.
In the simplest cases the neutrino spectra could be
completely degenerate at the neutrino sphere, or neutrino
self-interactions could produce flavor equilibration.

The reconstruction of the neutrino spectra and the
determination of the total gravitational binding energy
of the newly formed neutron star
has already been investigated in several works, such as
refs.~\cite{Minakata:2008nc,Lujan-Peschard:2014lta,An:2015jdp}. 
Of course, the answer to these issues depends on the constraints
imposed to the flux parameters in the likelihood analysis. 
Ref.\ \cite{Minakata:2008nc}  has studied in great detail the
electron antineutrino signal associated with inverse beta decay
in water Cherenkov detectors. A six degrees of freedom likelihood
analysis of the supernova signal has been performed considering not
only the total and the average energies of each neutrino species,
but also the pinching parameters. As an outcome, a continuous
degeneracy in the reconstructed parameters has been shown to exist,
which appears to preclude a precise determination of
the supernova neutrino fluxes. Ref.\ \cite{Lu:2016ipr}
shows that the explosion energy can be determined up to
$13 \%$ precision at $1 \sigma$ confidence level in JUNO
detector, assuming however that the pinching is fully known. In
ref.\ \cite{GalloRosso:2017hbp} we have shown that the
combined analysis of
events associated with inverse beta decay and electron scattering in
water Cherenkov detectors allows to determine the gravitational
binding energy of the neutron star with a precision of $11\%$ in a
detector like Super-Kamiokande, if a galactic supernova explodes.
The addition of neutral current neutrino-oxygen scattering
does not change the conclusion because of limited statistics.
Finally we have shown that a precise measurement of the
gravitational binding energy of the newly formed neutron star
allows to determine the star compactness within
$10 \%$ \cite{GalloRosso:2017hbp}. 

In this manuscript, we explore how precisely (all) the supernova
emission parameters can be reconstructed using data from water
Cherenkov detectors.  Differently from ref.\ 
\cite{GalloRosso:2017hbp} we consider Hyper-Kamiokande, 
that will very likely be built and that one can regard as 
the ultimate water Cherenkov detector.
The outcomes of this analysis are compared 
with those allowed by Super-Kamiokande,
that represents the best that supernova neutrino science can
achieve nowadays by means of water Cherenkov detectors.
The data set used for the analysis of Super-Kamiokande is the same
as the one in ref.\ \cite{GalloRosso:2017hbp} with the difference
that here we analyze all features of the neutrino spectra. 
We employ the well-motivated  
quasi-thermal neutrino spectra assuming that
flavor conversion is governed by the Mikheev-Smirnov-Wolfenstein
effect \cite{Wolfenstein:1977ue,Mikheev:1986gs}. 
We perform a likelihood analysis with 10 degrees of freedom
(9 for the flux parameters +1 for the $\nu$-oxygen cross
section uncertainty),
leaving in particular the parameters of the electron, muon and tau
neutrino fluxes unconstrained. 
The neutrino mass ordering is not yet known. We fix it to normal, as
template case, having in mind that combined analysis of oscillation
data show a slight preference for it, although statistical
significance is still low \cite{Capozzi:2017ipn}.
In the present analysis we combine events from inverse beta decay,
neutrino-electron elastic scattering and
neutral current interactions on oxygen nuclei. 
The attitude of our investigation is rather conservative, since, 
in particular, we do not assume any definite energy 
partition among the neutrino flavors {\em a priori}, as most 
commonly done in previous studies. Our choice is motivated
by the fact that the current supernova simulations, albeit not yet
definitive,  do not support the occurrence of an exact
equipartition, neither during the accretion
nor during the cooling phases of emission. Nonetheless, we
perform also another analysis, where the {\it ansatz} on energy
equipartition is implemented, to allow a comparison
with results available in the literature.

The paper is structured as follows. Section \ref{sec:StatAn}
presents the methods employed for the likelihood analysis of the 
events. The inputs of our calculations are described including
the neutrino fluences and the cross sections, and we compare
our outcomes with those of ref.\ \cite{Minakata:2008nc} when only
inverse-beta decay is considered and a 6 degrees of freedom
likelihood analysis performed. Section \ref{sec:risu} presents
the numerical results from the full 10-dimensional
likelihood analysis, and section \ref{sec:conclu} presents
our conclusions.

\section{Description of the statistical analysis}
\label{sec:StatAn}
\subsection{Detector properties}
\label{sec:DetectProp}

In our statistical analysis of a core-collapse
supernova explosion, we consider neutrino
events as can be seen by water Cherenkov detectors,
namely Super-Kamiokande %\cite{Fukuda:2002uc}
and Hyper-Kamiokande. %\cite{HK:2016dsw}. 
The former
is up to now the largest Cherenkov detector
in operation, with a total (fiducial) mass of
50 (22.5) kton of water \cite{Fukuda:2002uc}, while the latter
should be built in the future and will have a total
(fiducial) mass of 516 (374) kton \cite{HK:2016dsw}.
In this study, as active mass for the
two detectors, we consider \SI{22.5}{\kilo\ton}
and \SI{374}{\kilo\ton}
respectively.

Neutrinos in water can be detected thanks to several reactions
--- see e.g.\ \cite{Ikeda:2007sa}.
The most important one for our purposes, that will be included
in the analysis, are the following three:
\begin{enumerate}
	\item inverse beta decay (IBD):
	$\mathrm{p} + \overline{\nu}_{\mathrm{e}}
	\rightarrow \mathrm{n} + \mathrm{e}^+$;
	\item elastic scattering on electrons (ES):
	$\nu + \mathrm{e}^-
	\rightarrow \nu + \mathrm{e}^-$;
	\item neutral current scattering on oxygen
	(OS): $\nu + \text{\textsuperscript{16}O}
	\rightarrow \nu+\text{\textsuperscript{16}O}^*
	\rightarrow \nu + \gamma + \text{X}$.
\end{enumerate}
The IBD process provides the highest number of
events. The positron emission is, with good
approximation, isotropic \cite{Vogel:1999zy}
while the ES events spread out in a cone of about
$20^\circ$ \cite{Nakahata:1998pz} that points
towards the neutrino direction. This is crucial
for the ES tagging, because the directional
discrimination reduces the IBD background
to the portion of the solid angle in which it
overlaps to the ES signal (3\%).
Additional rejections can be performed applying an
energy cut --- 20\% on events with energy lower
than 30 MeV \cite{Pagliaroli:2009qy} --- and
neutron tagging. In this case, the rejection
power is 20\% for clear water
\cite{Pagliaroli:2009qy} and reaches
90\% with gadolinium-loading
\cite{Beacom:2003nk, Laha:2013hva}.

In this study we assume a 100\% tagging efficiency
on ES (and IBD) events above the detection threshold
of \SI{5}{\mega\electronvolt}, supposed to be the
same for Super-Kamiokande
\cite{Beacom:1998ya,Smy:2010zza}
and Hyper-Kamiokande \cite{HK:2016dsw}.
Concerning the OS signal, it is expected
to be within a window of $4\div
\SI{9}{\mega\electronvolt}$,
obtained combining the window covered by the expected
gamma lines ($\approx 5.3\div
\SI{7.3}{\mega\electronvolt}$)
\cite{Langanke:1995he} with
the energy resolution of the detector
($\approx 1.1\div \SI{1.3}{\mega\electronvolt}$)
\cite{Fukuda:2002uc}.
In this region, it cannot be disentangled from
the many more IBD and ES background events.
Thus, we are bound to consider the
sum of the three contributions. This constitutes
the neutral current low energy region (NCR)
that includes the OS signal.

\subsection{Choice of the parameters and of the initial values}
\label{sec:1:2}
We consider the neutrino signal as distributed
over a time integrated flux (fluence) and,
as usual practice, we assume
the four species $\nu_{\mu}$, $\overline{\nu}_{\mu}$,
$\nu_{\tau}$, $\overline{\nu}_{\tau}$, to be characterized
by the same emission spectrum. Therefore,
each of them is denoted with the same subscript,
$\nu_x$.
The fluence is described by a simple
analytical form that is in agreement with numerical
simulations and has been employed in refs.
\cite{Lujan-Peschard:2014lta,Tamborra:2012ac}.
This reads\footnote{In the following we put
$\hbar = c = k_B = 1$.}
\begin{equation}\label{eq:dFdE}
\mathcal{F}_i^0\left(E_{\nu}\right) = 
	\frac{\mathrm{d} F_i^0}{\mathrm{d} E_{\nu}} = 
	\frac{\mathcal{E}_i}{4\pi D^2}
	\frac{E_{\nu}^{\alpha_i}\:
	e^{-E_{\nu}/T_i}}{T_i^{\alpha_i +2 }\:
	\Gamma\left(\alpha_i+2\right)}
	\quad\text{with}\quad i = \nu_{\mathrm{e}},\,
	\overline{\nu}_{\mathrm{e}},\,\nu_x;
\end{equation}
where $E_{\nu}$ is the neutrino energy, $\Gamma(x)$
the Euler gamma function, $T_i$ the temperature
\begin{equation}
	T_i = \frac{\langle E_i\rangle}{(\alpha_i + 1)},
	\label{eq:TEmedia}
\end{equation}
and $\alpha_i$ the pinching parameter.

\begin{table}[t]
	\centering
	\begin{tabular}{|l|MMM|}
		\hline
			& \multicolumn{2}{c}{$\nu_{\mathrm{e}}$\Tstrut}
			& \multicolumn{2}{c}{
			$\overline{\nu}_{\mathrm{e}}$\Bstrut}
			& \multicolumn{2}{c|}{$\nu_x$}\\
			\hline
			$\mathcal{E}_i^*\:[10^{53}\:\mathrm{erg}]$\Tstrut
			& 0.5 & [0.2,\,1] & 0.5 & [0.2,\,1]
			& 0.5 & [0.2,\,1]\\
			$\langle E_i^* \rangle\:[\mathrm{MeV}]$
			& 9.5 & [5,\,30] & 12 & [5,\,30]
			& 15.6 & [5,\,30]\\
			$\alpha_i^*$ & 2.5 & [1.5,\,3.5] & 2.5 & [1.5,\,3.5]
			& 2.5 & [1.5,\,3.5]\Bstrut\\
			\hdashline
			$\kappa^*$ &
			\multicolumn{2}{c}{\Tstrut} &
			1 & [0.8,\,1.2] & \multicolumn{2}{c|}{\Bstrut}\\
			%\hline
			%$\varepsilon$ & \multicolumn{2}{c}{~} & 1 & [0.8,1.2]\\
			\hline
	\end{tabular}
	\caption{True parameter values assumed in the
	likelihood analysis and priors inside which they are
	free to vary (see text).
	The first three rows describe the astrophysical parameters,
	while the fourth concerns the uncertainty in the neutral
	current neutrino-oxygen  cross sections --- see
	eq.~\protect\eqref{eq:LikOS}. Note that the present analysis
	is based on
	the same hypothesis and priors as the one presented in
	ref.\ \cite{GalloRosso:2017hbp}.}
 	\label{tab:param}
\end{table}

The first set of parameters needed in
order to fully characterize the initial
emission is given by
the mean energies $\langle E_i\rangle$, univocally
connected to the temperatures $T_i$ by the pinching
parameter through eq.~\eqref{eq:TEmedia}.
We choose as reference values the quantities
reported in ref.\ \cite{Lujan-Peschard:2014lta},
in full agreement with SN 1987A
and summarized in table~\ref{tab:param}.

The quantity $\alpha$ takes into account the deviation of the tails
from the thermal
distribution. The exact value of this parameter is far
from being  exactly known.\footnote{The case $\alpha = 2$
identifies a Maxwell-Boltzmann
distribution, while $\alpha = 2.30$ closely
approximates a Fermi-Dirac distribution
(for fixed mean energy).}
Comparing the results
from several models $\alpha$ is found to vary
between 2.5 and 5 for the time dependent flux 
\cite{Keil:2002in}. Note that the combination of
detection channels with different energy thresholds such as 
one- and two-neutron emission in a lead-based observatory
like HALO can be used to reduce significantly the pinching
parameter range \cite{Vaananen:2011bf}.
The fluence eq.~\eqref{eq:dFdE} is observed to be closer to a thermal distribution than the time-dependent flux. A reasonable conservative
interval is \cite{Vissani:2014doa}
\begin{equation}
\label{eq:rangeVissani}
	1.5 \le \alpha\le 3.5.
\end{equation}
For the analysis of simulated
data we select the true values $\alpha_i^*$ to be equal
for all the tree species and in the middle of this
range (see table~\ref{tab:param}).
Such a value is supported by the most recent simulations
and does not contradict the data from supernova
1987A.

In our simulations the true value of the total energy
emitted in neutrinos is assumed to be
$\mathcal{E}^* = \SI{3e53}{\erg}$ at a
distance $D=\SI{10}{\kilo\parsec}$, the latter
supposed to be known from astronomical observations
with negligible error. This is a reasonable value and
can be justified in many ways
\cite{Costantini:2005un,Mirizzi:2006xx,Adams:2013ana}.
The value of $\mathcal{E}^*$ is the one typically
used in the literature (see e.g.\ 
\cite{Lujan-Peschard:2014lta,Loredo:2001rx,
Pagliaroli:2008ur}), together with the equipartition hypothesis,
namely the assumption that the initial energy is
equally divided among each neutrino flavor
\begin{equation}
	\label{eq:svero}
	\mathcal{E}^*_i = \SI{5e52}{\erg}
	\quad\text{where}\quad i =
	\nu_{\mathrm{e}},\, \overline{\nu}_{\mathrm{e}},\,
	\nu_{x}.
\end{equation}
This is a useful simplification
although not very reliable,
as it is known to be true only within a factor of two
\cite{Keil:2003sw,Raffelt:2005fb}. For this reason,
the equipartition hypothesis is taken only to fix
the initial value of the total emitted energies
\eqref{eq:svero}, while in the likelihood analysis
we do not impose any constraint unless 
explicitly stated.

In our analysis we consider that 
neutrinos change their flavor
from the neutrinosphere to the detector according to
the Mikheyev-Smirnov-Wolfenstein (MSW)
effect \cite{Wolfenstein:1977ue,Mikheev:1986gs}.
In this case, and for normal neutrino mass ordering that we take as reference example,
 the fluences become
\cite{Dighe:1999bi}
\begin{equation}\label{eq:fluIBD}
	\begin{cases}
		\mathcal{F}_{\Pnue} &=
		\mathcal{F}_{\Pnux}^0\\
		\mathcal{F}_{\APnue} &= 
		P_{\mathrm{e}}
		\cdot \mathcal{F}_{\APnue}^0 + 
		(1-P_{\mathrm{e}}) \cdot\mathcal{F}_{\Pnux}^0,
	\end{cases}
\end{equation}
for electron neutrinos and antineutrinos.
The superscript 0 refers to the
distributions at the neutrinosphere, while
$P_{\mathrm{e}}$ is the survival probability
\cite{Capozzi:2017ipn}
\begin{equation}
	P_{\mathrm{e}} = \left| U_{e1} \right|^2 =
	\left| \cos\theta_{12} \cos\theta_{13} \right|^2
	\approx 0.70.
\end{equation}
It is useful to remind that in the case of normal mass
ordering the $\nu_x$  at the neutrinosphere are measured as
$\nu_{\mathrm{e}}$.

\subsection{Events extraction}
\label{sec:1:3}

We give here a detailed description of the cross
sections, of the expected distributions and of the number
of extracted events for the processes of interest. Such events
are summarized in table \ref{tab:estratti},
for Super-Kamiokande\footnote{Note
	that the numbers reported for Super-Kamiokande are
	the same as the ones reported in table~2 of
	ref.\ \cite{GalloRosso:2017hbp}, since we use here the same
	set of extracted data for Super-Kamiokande as in ref.\ 
	\cite{GalloRosso:2017hbp}.} as well as Hyper-Kamiokande.

\subsubsection{Inverse Beta Decay}

Following the calculation reported in
\cite{strumia-vissani,Vissani:2014doa},
the IBD cross section can be written as
\begin{equation}
	\sigma_{\mathrm{IBD}} = \kappa_{\mathrm{IBD}}\;
	p_e E_e \quad\text{with}\quad E_e = E_{\nu} - \Delta,
\end{equation}
where $E_{\nu}$ is the energy of the
incoming neutrino, $\Delta
\approx\SI{1.293}{\mega\electronvolt}$,
$E_e$ and $p_e$ are the positron
energy and momentum. The coefficient
$\kappa_{\mathrm{IBD}}$ is parametrized as
\begin{equation}
	\kappa_{\mathrm{IBD}} = e^{-0.07056 x + 0.02018 x^2
	- 0.001953 x^4} \times\SI{e-43}{\centi\meter\squared}
	\quad\text{where}\quad
	x=\log\left(\frac{E_{\nu}}{\mathrm{MeV}}\right).
\end{equation}
The differential rate of events is,
with good approximation, given by
\begin{equation}
	\mathcal{N}_{\mathrm{IBD}}\left(E_e\right) \equiv
	\frac{\mathrm{d}{\text{N}}_{\mathrm{IBD}}}{\mathrm{d}{E_e}}
	\approx
	N_p\, \mathcal{F}_{\APnue}
	\left(E_{\nu}\right)
	\sigma_{\mathrm{IBD}}\left(E_{\nu}\right)
	J\left(E_{\nu}\right),
	\label{eq:dNdEI}
\end{equation}
with $N_p$ number of target protons, 
$\mathcal{F}_{\APnue}$ from
eq.~\eqref{eq:fluIBD} and $J\left(E_{\nu}\right)$
\begin{equation}
	J\left(E_{\nu}\right) = \frac{\left(1+E_{\nu}/m_p\right)^2}
	{1+ \Delta/m_p}\quad\text{with}\quad
	E_{\nu} = \frac{E_e + \Delta}{1 - E_e / m_p},
\end{equation}
$m_p$ being the proton mass.

\begin{table}[tp]
	\centering
	\begin{tabular}{|r|cccc|c|cccc|c|}
		%\hline
		\cline{2-11}
		\multicolumn{1}{c|}{~}
		& \multicolumn{5}{c|}{\Tstrut Super-Kamiokande \Bstrut}
		& \multicolumn{5}{c|}{\Tstrut Hyper-Kamiokande \Bstrut}\\
		\cline{2-11}
		\multicolumn{1}{c|}{~}
 		& $\nu_{\mathrm{e}}$ & $\overline{\nu}_{\mathrm{e}}$
 		& $4\nu_{x}$
 		\Bstrut & \Tstrut sum & $n_j$
 		& $\nu_{\mathrm{e}}$ & $\overline{\nu}_{\mathrm{e}}$
 		& $4\nu_{x}$
 		\Bstrut & \Tstrut sum & $n_j$\\
 		\hline
 		IBD & --- & 2900 & 1672 & 4572 \Tstrut & 4565
 		% HK
 		& --- & 48198 & 27795 & 75993 & 76427\\
 		ES \Bstrut
 					& 14.7	& 24.7	& 187	& 226	& 237
 		% HK
 					& 244 	& 410	& 3105	& 3759	& 3848\\
 		\hdashline
 		NCR\Tstrut 	& 12		& 345	& 204	& 561	& 554
 		% HK
 					& 192 	& 5732	& 3392	& 9316	& 9210\\
 		%\hdashline
 		OS sig.\   	& 0.53	& 2.04 	& 43.0 	& 45.5 &	
 				   	& 9		& 34		& 714	& 757 &\\
 		IBD bkg.\  	& ---	& 324 	& 77.8 	& 401 &
 				   	& ---	& 5379	& 1293	& 6672 &\\
 		ES	bkg.\  	& 11.0 	& 19.2 	& 83.3 	& 114 &
 				   	& 183	& 319	& 1385	& 1887 &\\
 		\hline
	\end{tabular}
	\caption{Number of expected events $\mathrm{N}^*$
	for IBD (first), ES (second) and
	NCR (third lines) in Super-Kamiokande
	(left) and Hyper-Kamiokande (right).
	In both cases the first three columns show the contributions
	for each neutrino flavor as well as their sum.
	The fourth columns on both sides refer to
	the values $n_j$ %($j = e,x$)
	extracted in our statistical analysis.
	The last three lines correspond to the neutral current
	neutrino-oxygen events and the IBD and ES background
	events in the low energy region
	($\approx 4 \div\SI{9}{\mega\electronvolt}$).}
 	\label{tab:estratti}
\end{table}

\begin{figure}[t]
\centering %scrivo
\subfloat[IBD in Super-Kamiokande]{\label{fig:skibd}
\includegraphics[width=0.48\textwidth]{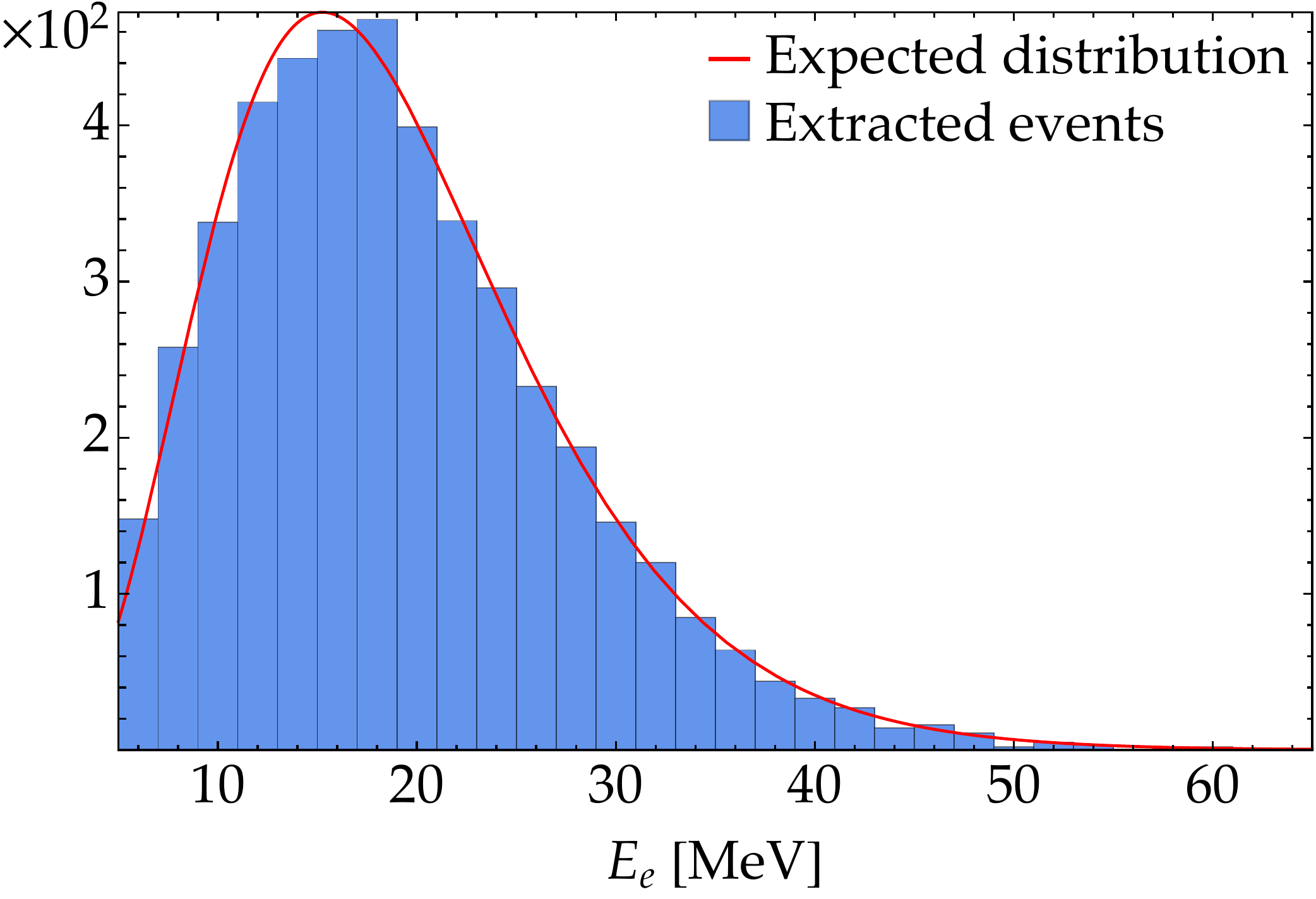}}
\subfloat[IBD in Hyper-Kamiokande]{\label{fig:hkibd}
\includegraphics[width=0.48\textwidth]{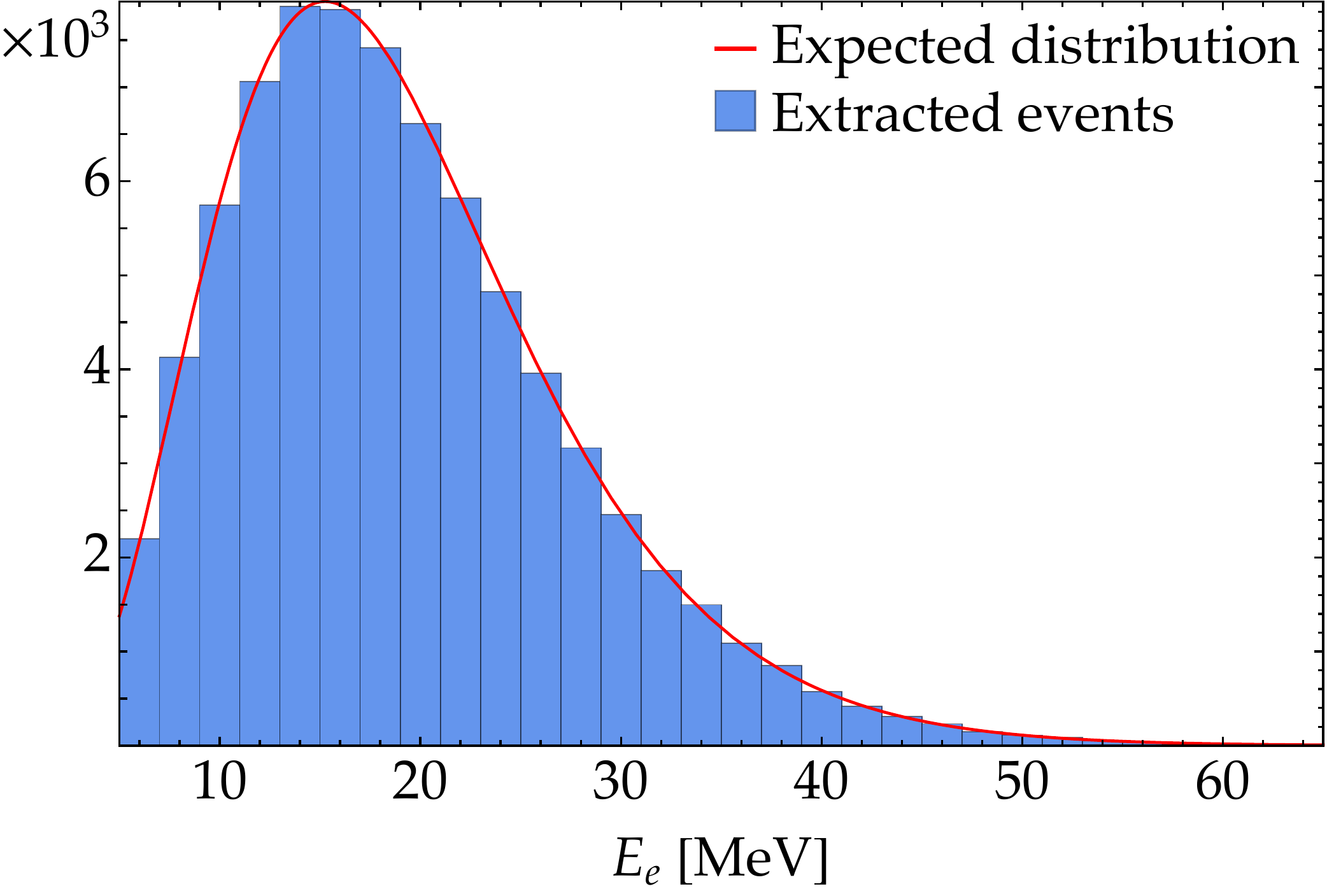}}\\
\subfloat[ES in Super-Kamiokande]{\label{fig:skes}
\includegraphics[width=0.48\textwidth]{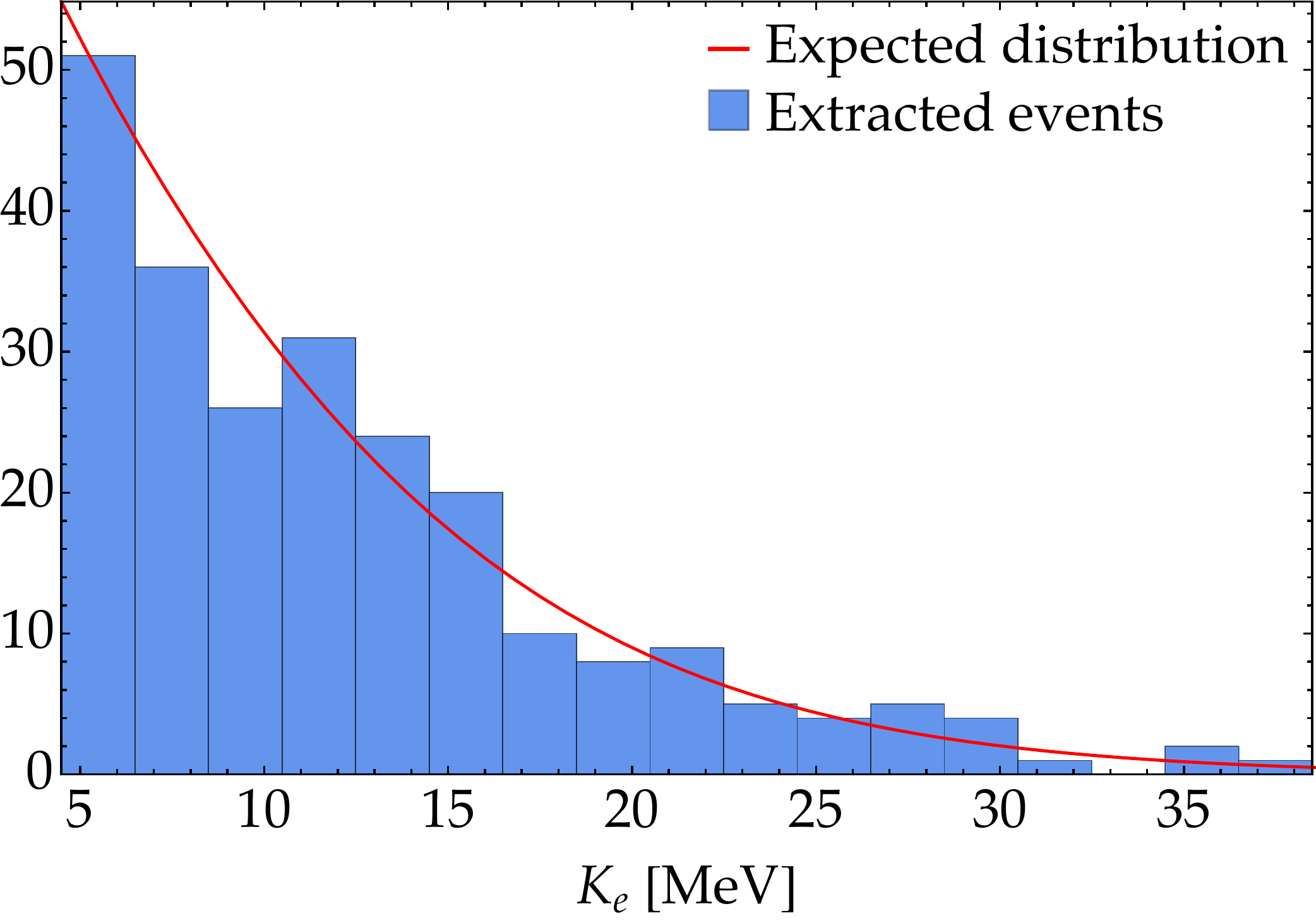}}
\subfloat[ES in Hyper-Kamiokande]{\label{fig:hkes}
\includegraphics[width=0.48\textwidth,height=0.3355\textwidth]{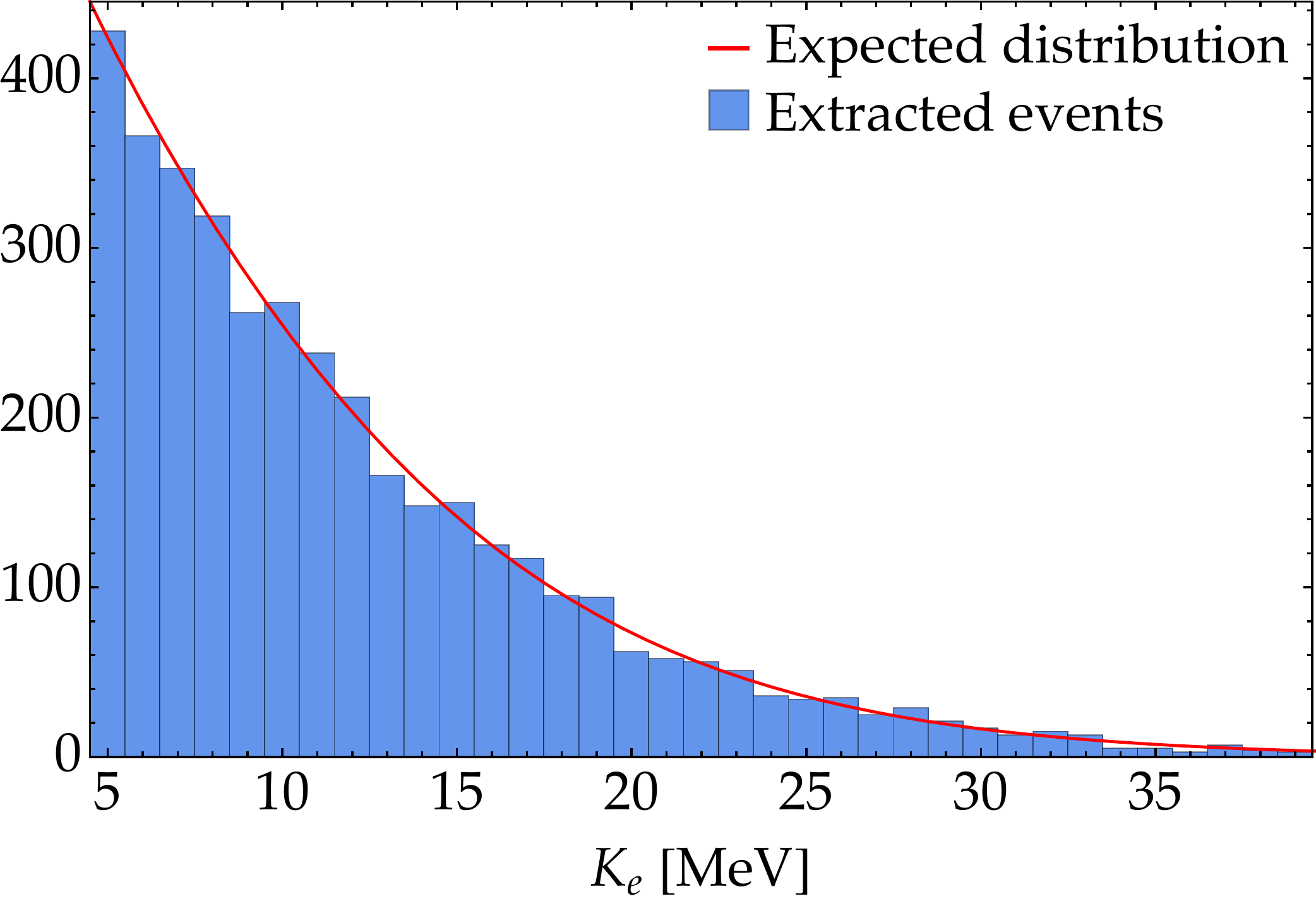}}
\caption{Extracted IBD (first row) and ES
(second row) events used for the analysis in
Super-Kamiokande (left column) and Hyper-Kamiokande
(right column). The expected distributions, obtained
by using eqs.~\protect\eqref{eq:dNdEI} and \protect\eqref{eq:dNES}
respectively, are plotted over the
corresponding histograms.}
\label{fig:estratti}
\end{figure}

The number of expected IBD
events for a neutrino emission characterized
by the parameters given in table~\ref{tab:param}
is obtained integrating eq.~\eqref{eq:dNdEI}.
We assume a threshold
of \SI{5}{\mega\electronvolt} both for Super-Kamiokande
\cite{Beacom:1998ya,Smy:2010zza} and Hyper-Kamiokande
together with 100\% efficiency.\footnote{
This assumptions is adopted to estimate the
ultimate sensitivity of water Cherenkov detectors.}
The number of expected events from the IBD process,
$\mathrm{N}_{\mathrm{IBD}}^*$, is reported in the
first row of table~\ref{tab:estratti}.
Given its value, the number of events analyzed
$n_{\mathrm{IBD}}$, is extracted
from a Poissonian distribution with average value
$\mathrm{N}_{\mathrm{IBD}}^*$. Each one of the
extracted events is characterized by a positron energy
$E_e$, randomly distributed according to
the spectrum \eqref{eq:dNdEI} and plotted
in figure~\ref{fig:skibd} 
for Super-Kamiokande and \ref{fig:hkibd} for Hyper-Kamiokande.

\subsubsection{Elastic scattering on electrons}

Concerning the elastic scattering on electrons,
the well-known tree level expression of the
cross section --- see e.g.\ refs.\ \cite{jb}
or \cite{Formaggio:2013kya} --- is
\begin{equation}
	\frac{\mathrm{d}\sigma_{\mathrm{ES},j}}
	{\mathrm{d}y} = \frac{2 G_F^2}{\pi}
	m_e E_{\nu} \left[g_{j}^2 + g_{j}^{\prime 2}
		\left(1-y\right)^2
		- g_{j} g_{j}' \frac{m_e}{E_{\nu}} y
	\right]\quad\mathrm{with}\quad
	j = \nu_{\mathrm{e}},\,
	\overline{\nu}_{\mathrm{e}},\,
	\nu_{\mu,\tau},\,\overline{\nu}_{\mu,\tau};
	\label{eq:SigmaES}
\end{equation}
where $G_F$ is the Fermi constant, $m_e$
is the electron mass, $E_{\nu}$ the neutrino energy and
$y$ is the electron kinetic energy $K_e = E_e - m_e$
over the neutrino energy
\begin{equation}
	0 < y = \frac{K_e}{E_{\nu}} < \left[
	1 + \frac{m_e}{2 E_{\nu}}\right]^{-1} .
\end{equation}
The various a-dimensional constants are listed
in table~\ref{tab:cost1}.

By multiplying 
the cross section by the fluence \eqref{eq:fluIBD}
$\mathcal{F}_i\left(E_{\nu}\right)$ for
each species and by the number $N_e$ of target electrons
and by performing an integration, we end up with an
analytical expression for the expected spectrum,
differential in the 
kinetic energy of the recoiling electron
\begin{equation}
\begin{aligned}\label{eq:dNES}
	\mathscr{N}_{\text{ES}}\left(K_e\right)\equiv
	\frac{\text{dN}_{\text{ES}}}{\mathrm{d}{K_e}}=
	\sum_{\nu = \nu_{\mathrm{e}},
	\overline{\nu}_{\mathrm{e}}, \nu_x}
	&\xi_{\nu}
	\left[
	\left(G_{1 \nu} + G_{2 \nu}\right)
	\mathscr{F}_{\nu}^{(+1)}-2\frac{K_e}{T_{\nu}}
	G_{2\nu}\mathscr{F}_{\nu}^{(0)} \right.
	\\
	&\left.
	+\frac{K_e}{T_{\nu}}\left(\frac{K_e}{T_{\nu}}
	G_{2\nu} - \frac{m_e}{T_{\nu}}G_{3\nu}\right)
	\mathscr{F}_{\nu}^{(-1)}\right],
\end{aligned}
\end{equation}
where $T_{\nu}$ is the same as \eqref{eq:TEmedia} and
the quantity $\xi_{\nu}$ is given by
\begin{equation}
	\xi_{\nu} = %\left(\hbar c\right)^2
	\frac{2 G_F^2 m_e}{\pi} 
	\frac{\mathcal{E}_{\nu}}
	{\langle E_{\nu}\rangle}
	\frac{N_e}{4\pi D^2}.
\end{equation}
Functions $\mathscr{F}_{\nu}^{(i)}\left(K_e\right)$
are defined as follows
\begin{equation}
	\mathscr{F}_{\nu}^{(i)}\left(K_e\right)=
	\frac{\Gamma \left(\alpha _{\nu }+i,
	(\zeta +1) K_e/T_{\nu }\right)}
	{\Gamma  \left(\alpha _{\nu }+1\right)}
	\quad\text{with}\quad\zeta =
	\frac{m_e / K_e}{1 +\sqrt{1 + 2 m_e / K_e}},
\end{equation}
where $\Gamma\left(a, z\right)$
is the incomplete gamma function
\begin{equation}
	\Gamma (a,z)=\int_z^{\infty} e^{-t}\; t^{a-1} \;
	\mathrm{d}t.
\end{equation}
Finally, the values of the numerical
constants $G_{j\nu}$ are listed in
table~\ref{tab:cost2}.

Integrating eq.~\eqref{eq:dNES} from the threshold of  
$K_{e,\mathrm{thr}} = \SI{5}{\mega\electronvolt} - m_e$ that is 
identical for Super-Kamiokande and Hyper-Kamiokande,
we obtain the number of expected ES
events, $\mathrm{N}^*_{\mathrm{ES}}$,
for the priors shown in 
table~\ref{tab:param}.
The expected quantities are reported in the second
row of table~\ref{tab:estratti}, together with the
randomly extracted number of events considered
in the present analysis.
Each one of the ES events is characterized by a
recoiling energy $K_e$, randomly distributed
according to  \eqref{eq:dNES}. The corresponding event
distributions are plotted
in figures~\ref{fig:skes} and \ref{fig:hkes}
for Super-Kamiokande and Hyper-Kamiokande respectively.

\begin{table}[t]
  \centering
  \subfloat[Constants in \protect\eqref{eq:SigmaES}]{%
    \begin{tabular}{|C|CC|}
		\hline
		& g \Tstrut & g' \Bstrut \\
		\hline
		\nu_{\mathrm{e}}\Tstrut & \frac{1}{2}+\sin^2\theta_W & \sin^2\theta_W \\
		\overline{\nu}_{\mathrm{e}} &  \sin^2\theta_W & \frac{1}{2}+\sin^2\theta_W	\\
		\nu_{\mu,\tau} & -\frac{1}{2}+\sin^2\theta_W	 &\sin^2\theta_W \\
		\overline{\nu}_{\mu,\tau}\Bstrut 
		&\sin^2\theta_W &-\frac{1}{2}+\sin^2\theta_W \\
		\hline
		\end{tabular}\label{tab:cost1}
  }
  \hspace{1cm}
  \subfloat[Constants in \protect\eqref{eq:dNES}]{%
    \begin{tabular}{|r|ccc|}
        \hline
        $G_{j\nu}$ &	1 & 2 & 3 \Tstrut\Bstrut\\
        \hline
        $\nu_{\mathrm{e}}$  & 0.073 & 0.054
        & -0.062 \Tstrut\\
		$\overline{\nu}_{\mathrm{e}}$ & 0.054 & 0.394
		& 0.099 \\
		$\nu_x$& 0.713 & 0.393 & 0.053\Bstrut \\
        \hline
		\end{tabular}\label{tab:cost2}
		}%
    \caption{Numerical constants used in
    eqs.~\protect\eqref{eq:SigmaES}
    (left table) and \protect\eqref{eq:dNES} (right table).
	The parameter $\theta_W$ is the Weinberg
	weak-mixing angle.}
\end{table}

\subsubsection{Neutral current neutrino-oxygen scattering}
\label{sec:HP:NOxyEv}

In order to compute the number of events due to
neutrino interactions on \textsuperscript{16}O
followed by gamma emission we take for the cross section
the parametrization 
\begin{equation}
	\sigma_{\mathrm{OS}}(E_{\nu}) \approx \sigma_0
	\left(E_{\nu}/\si{\mega\electronvolt} - 15\right)^4
	\vartheta\left(E_{\nu}/\si{\mega\electronvolt} - 15\right),
	\label{eq:CsecParam}
\end{equation}
where $E_{\nu}$ is the energy of the
incoming neutrino and $\vartheta$ the Heaviside
theta function. The coefficient $\sigma_0$ is different
for neutrinos or antineutrinos. We will refer to it as 
 $\sigma_{0\nu}$  and $\sigma_{0\bar{\nu}}$ respectively. 
 The expression (\ref{eq:CsecParam})
 has proposed by ref.\ \cite{Beacom:1998ya}
 to parametrize the cross sections computed with
 a microscopic approach.
 
In order to determine the value of $\sigma_0$,
we reproduce the flux-averaged cross sections of
ref.\ \cite{Langanke:1995he} (table~I), shown in table 	\ref{tab:CsecFold}. These have been obtained by folding the numerically computed cross sections $\sigma^*$  
\begin{equation}
	\sigma_{tot} = \int_0^{\infty}
	f_{FD}\left(E_{\nu} \middle| T,\,\mu \right) \sigma^*(E_{\nu})
	\mathrm{d} {E_\nu}.
	\label{eq:sigmaTot}
\end{equation}
with a pinched Fermi-Dirac spectrum $f_{FD}$
\begin{equation}
	f_{FD}\left(E\right) = 
	\left[T^3\,\tilde{F}_2\left(\frac{\mu}{T}
	\right)\right]^{-1}
	%\frac{1}{T^3\,F_2\left(\frac{\mu}{T}\right)}
	\frac{E^2}{1 + \exp\left[\left(E-
	\mu\right)/T\right]},
\end{equation}
where $\tilde{F}_n\left(\eta\right)$
is the Fermi integral of order $n$
\begin{equation}\label{eq:Fermi}
	\tilde{F}_n(\eta) = \int_0^\infty
	\frac{x^n \mathrm{d} x}{1+ \exp\left(x-
	\eta\right)}.
\end{equation}
For each process shown in the table and spectrum, we describe 
the numerical cross sections $\sigma^*$ 
with the expression given in eq.~\eqref{eq:CsecParam}. 
We carry out two calculations with two different neutrino spectra referred to as
$f_{FD,1}$ and $f_{FD,2}$. For the first case,  the temperature is
$T = \SI{8}{\mega\electronvolt}$
and the chemical potential
$\mu = 0$;  while the second the spectrum
is characterized by  $T = \SI{6.26}{\mega\electronvolt}$ and $\mu = 3 T$.
\begin{equation}
	f_{FD,1}
	\begin{cases}
		\sigma_{0\nu,1} = \SI{4.08E-22}{\femto\meter\squared}\\
		\sigma_{0\bar{\nu},1} = \SI{3.14E-22}{\femto\meter\squared}\\
	\end{cases}\qquad
	f_{FD,2}
	\begin{cases}
		\sigma_{0\nu,2} = \SI{4.33E-22}{\femto\meter\squared}\\
		\sigma_{0\bar{\nu},2} = \SI{3.52E-22}{\femto\meter\squared}\\
	\end{cases}.
\end{equation}
Since for each spectrum  we have two different values,
the parameters taken as reference are a mean between
$\sigma_{0\nu,1}$ and $\sigma_{0\nu,2}$ or between
$\sigma_{0\bar{\nu},1}$ and $\sigma_{0\bar{\nu},2}$
\begin{equation}
	\sigma_{0\nu}=\SI{4.21E-22}{\femto\meter\squared}
	\quad\text{and}\quad
	\sigma_{0,\bar{\nu}} =
	\SI{3.33E-22}{\femto\meter\squared}.
\end{equation}

\begin{table}[tbp]
		\begin{center}
			\begin{tabular}{|r|cr|cr|}
				\hline
				& \multicolumn{2}{c|}{Neutrinos} 
				& \multicolumn{2}{c|}{Antineutrinos} \\
				\hline
				& \multirow{2}{*}{Reaction}
				& \multicolumn{1}{c|}{$\sigma_{tot}$}
				& \multirow{2}{*}{Reaction}
				& \multicolumn{1}{c|}{$\sigma_{tot}$}\\
			& & \multicolumn{1}{c|}{$(10^{-16} \si{\femto\meter\squared})$} \Bstrut
			& & \multicolumn{1}{c|}{$(10^{-16} \si{\femto\meter\squared})$}\\
				\hline
			\multirow{2}{*}{$f_{FD,1}$}
			&$\mathrm{\textsuperscript{16}O}
			(\nu, \nu'\mathrm{p}\gamma)
			\mathrm{\textsuperscript{15}N}$
			& 1.41 \Tstrut
			&$\mathrm{\textsuperscript{16}O}
			(\bar{\nu}, \bar{\nu}'\mathrm{p}\gamma)
			\mathrm{\textsuperscript{15}N}$
			& 1.09 \\
			&$\mathrm{\textsuperscript{16}O}
			(\nu, \nu'\mathrm{n}\gamma)
			\mathrm{\textsuperscript{15}O}$
			& 0.37 \Bstrut
			&$\mathrm{\textsuperscript{16}O}
			(\bar{\nu}, \bar{\nu}'\mathrm{n}\gamma)
			\mathrm{\textsuperscript{15}O}$
			& 0.28\\
			\hline
			\multirow{2}{*}{$f_{FD,2}$}
			&$\mathrm{\textsuperscript{16}O}
			(\nu, \nu'\mathrm{p}\gamma)
			\mathrm{\textsuperscript{15}N}$
			&0.72 \Tstrut
			&$\mathrm{\textsuperscript{16}O}
			(\bar{\nu}, \bar{\nu}'\mathrm{p}\gamma)
			\mathrm{\textsuperscript{15}N}$
			& 0.59 \\
			&$\mathrm{\textsuperscript{16}O}
			(\nu, \nu'\mathrm{n}\gamma)
			\mathrm{\textsuperscript{15}O}$
			& 0.18 \Bstrut
			&$\mathrm{\textsuperscript{16}O}
			(\bar{\nu}, \bar{\nu}'\mathrm{n}\gamma)
			\mathrm{\textsuperscript{15}O}$
			& 0.14\\
				\hline
			\end{tabular}
		\end{center}
		\caption{Flux-averaged 
		$\nu$-\textsuperscript{16}O cross sections 		
		eq.~\protect\eqref{eq:sigmaTot} followed by $\gamma$
		emission (from table I of ref.\ 
		\protect\cite{Langanke:1995he}). These are obtained 
		by folding the numerical
		cross sections $\sigma^*$ with two different pinched
		Fermi-Dirac spectra. The first one,
		$f_{FD,1}$, has temperature
		$T = \SI{8}{\mega\electronvolt}$
		and chemical potential $\mu = 0$. The second
		one, $f_{FD,2}$, is characterized by
		$T = \SI{6.26}{\mega\electronvolt}$
		and $\mu = 3 T$. Note that the cross
		sections are computed using a microscopic calculation.}
		\label{tab:CsecFold}
	\end{table}

The analytical expression is known to have an uncertainty of
$\sim 10\%$ \cite{Beacom:1998ya} that we have to
take into account in the likelihood analysis.
In order to do so, we introduce a tenth
parameter, $\kappa$, as a multiplicative constant
for the whole cross section (identical for neutrinos
and antineutrinos). The final expression becomes
\begin{equation}	\label{eq:cSecNC2}
	\sigma_{\mathrm{OS}}\left(E_{\nu}\right) = 
		\kappa\,
		\sigma_{0,\nu}\left(E_{\nu}/
		\si{\mega\electronvolt} -15 \right)^4
		\vartheta\left(E_{\nu}/
		\si{\mega\electronvolt} - 15\right)
\end{equation}
for neutrinos and similarly for antineutrinos
(by replacing $\sigma_{0,\nu}$ with $\sigma_{0,\bar{\nu}}$).
We assume the true value of $\kappa$ to be
\begin{equation}
\label{eq:epsivero}
\kappa^* = 1
\end{equation}
and vary it according to a gaussian distribution
of mean 1 and variance
\begin{equation}
	\sigma_{\kappa} = 0.1.
\end{equation}
This uncertainty reflects the $10\%$ precision
on the cross section hopefully reachable in the
future. Although it may seem quite optimistic in
comparison with the available theoretical predictions
in the literature (see e.g.\ refs.\ 
\cite{Haxton:1987kc,Kolbe:1992xu,Langanke:1995he,Kolbe:2002gk})
we anticipate that
the results depend only weakly on the inclusion
of the NCR reactions. Note that even if the numerical
neutrino-oxygen
cross sections were employed, one should nevertheless include the
corresponding current theoretical uncertainty.
Moreover, as discussed in ref.\ \cite{GalloRosso:2017hbp},
variations of $\kappa$ and $\sigma_\kappa$ parameters 
lead only to negligible changes.

In order to obtain the number of expected OS events, 
we have multiplied  eq.~\eqref{eq:cSecNC2}
by the fluence $\mathcal{F}_i\left(E_{\nu}\right)$
and the number $N_{\rm O}$ of oxygen nuclei.
Then, integrating on the neutrino energy,
we end up with the number of expected OS events
for the parameters described in table~\ref{tab:param}. Results are given in table~\ref{tab:estratti} for both Super-Kamiokande
and Hyper-Kamiokande. We recall that
the quantity used in our analysis is the total
amount of events in the NCR region, obtained
by adding  the IBD and ES background  events with the OS signal events in the
OS signal window (see section~\ref{sec:DetectProp}).
The expected number of NCR events as well as the
random quantities extracted for our analysis  are also shown
in table~\ref{tab:estratti}.

\subsection{Likelihoods}
\label{sec:1:4}

\begin{table}[t]
	\centering
	\begin{tabular}{|ccl|}
		\hline
		Range [MeV] & $\Delta
		\left.E_{e}\right|_{\text{bin}}$ [MeV] &
		\Tstrut\Bstrut \\
		\hline
		$5  \le E_e \le 40$ & 0.5 \Tstrut &
		$\Rightarrow$ 70 bins\\
		$40 \le E_e \le 50$ & 1   &
		$\Rightarrow$ 10 bins\\
		$50 \le E_e \le 56$ & 2   &
		$\Rightarrow$ 3 bins\\
		$56 \le E_e \le 60$ & 4   &
		$\Rightarrow$ 1 bin\\
		$60 \le E_e \le 100$& 40  &
		$\Rightarrow$ 1 bin \Bstrut\\
		\hline
	\end{tabular}
	\caption{Bin widths (second column) used in IBD
	and ES likelihoods as a function of positron/electron energy
	ranges (first column). The number of bins used in each energy
	range is also shown (third column).}
	\label{tab:binLik}
\end{table}

In order to analyze the IBD and ES extracted
events, we use a binned likelihood
\begin{equation}
\label{eq:LikBin}
	\mathcal{L}_{j}\left(\text{param.}\right)\propto
	\prod_{i=1}^{N_{\text{bin}}}
	\frac{\nu_i^{n_i}}{n_i} e^{-\nu_i}
	\quad\text{with}\:j=\text{IBD, ES};
\end{equation}
where $\nu_i$ is the number of expected
events for the process $j$
(as a function of parameters) in the
$i$-th bin and $n_i$ is the number observed
in the simulation, in the same bin.
Bin widths are not uniform, but vary
according to table~\ref{tab:binLik}, so as
to preserve the relevant features of the distributions.
The same choice is adopted in
refs.\ \cite{Minakata:2008nc,GalloRosso:2017hbp}. 

Concerning the neutral current events on oxygen,
the likelihood form would be in
principle simply Poissonian, since we have to
compare the number of expected events
$\text{N}_{\mathrm{NCR}}$ (as a function
of the parameters) with the extracted one
$n_{\mathrm{NCR}}$.\footnote{Since the number of
events is of the order of $\sim 500$
for Super-Kamiokande and $\sim 10^4$ for
Hyper-Kamiokande, we can also replace the
Poissonian functional form with the (simpler)
gaussian one.} Nevertheless, we have also
to take into account the uncertainty on the
cross section via the $\kappa$ parameter
eq.~(\ref{eq:cSecNC2}--\ref{eq:epsivero}).
Therefore, the likelihood expression becomes
\begin{equation}
\label{eq:LikOS}
	\mathcal{L}_{\mathrm{NCR}}
	\left(\text{param.}\right) =
	\left(2\pi \text{N}_{\mathrm{NCR}}
	2\pi \sigma_\kappa^2
	\right)^{-1/2}
	\exp\left[
	-\frac{\left(n_{\mathrm{NCR}} -
	\text{N}_{\mathrm{NCR}}\right)^2}{2
	\text{N}_{\mathrm{NCR}}} -
	\frac{\left(\kappa -1 \right)^2}
	{2\sigma_\kappa^2}
	\right].
\end{equation}

In order to quantify the relevance of the various
reactions in the reconstruction of the emission
parameters, we apply the same strategy as 
in ref.\ \cite{GalloRosso:2017hbp} and perform three different
analyses, for Super-Kamiokande and Hyper-Kamiokande
separately.  First we perform an analysis with IBD events alone. 
Then a second one where we add the ES information. Finally the
third one exploits IBD, ES and NCR. In the three analysis we assume
100\% tagging efficiency for IBD and ES
processes above \SI{5}{\mega\electronvolt}
(see section~\ref{sec:DetectProp}).
Thus, for each analysis the global likelihood is
expressed as the product of the likelihoods of
the single processes.

As in ref.\ \cite{GalloRosso:2017hbp}, a
Monte Carlo approach is employed:
each variable is free to vary inside the corresponding
prior (table~\ref{tab:param}). Then, the
$n-$dimensional region described by these priors
is covered with random points $P$. Each of them
is accepted within a certain confidence level (CL)
if its likelihood satisfies the relation
\begin{equation}
	\log\mathcal{L}\left(P\right)\ge
	\log\mathcal{L}_{max} - \frac{A}{2},
	%\quad\text{with}\quad
	%\int_0^{A} \chi^2(N_{\mathrm{dof}}; z)
	%\,\mathrm{d}{z} = \mathrm{CL},
\end{equation}
where $\mathcal{L}_{max}$ is the likelihood
maximum inside the prior and $A$
is defined with an integral of a
chi-square distribution with
$N_{\mathrm{dof}}$ degrees of freedom
\begin{equation}
\int_0^{A} \chi^2(N_{\mathrm{dof}}; z)
	\,\mathrm{d}{z} = \mathrm{CL},
\end{equation}
For instance, considering $\mathrm{CL} = 0.9973$, namely
$3\sigma$, we find $A = 11.8292,\, 18.2053,\,26.9011$
for $N_{\mathrm{dof}} = 2,\,5,\,10$ respectively.

\begin{figure}[t]
	\centering
	\subfloat[$\langle E_{\APnue}
	\rangle$--$\langle E_{\Pnux}\rangle$ projection]
	{\label{fig:mik1}
	\includegraphics[width=0.33\textwidth]{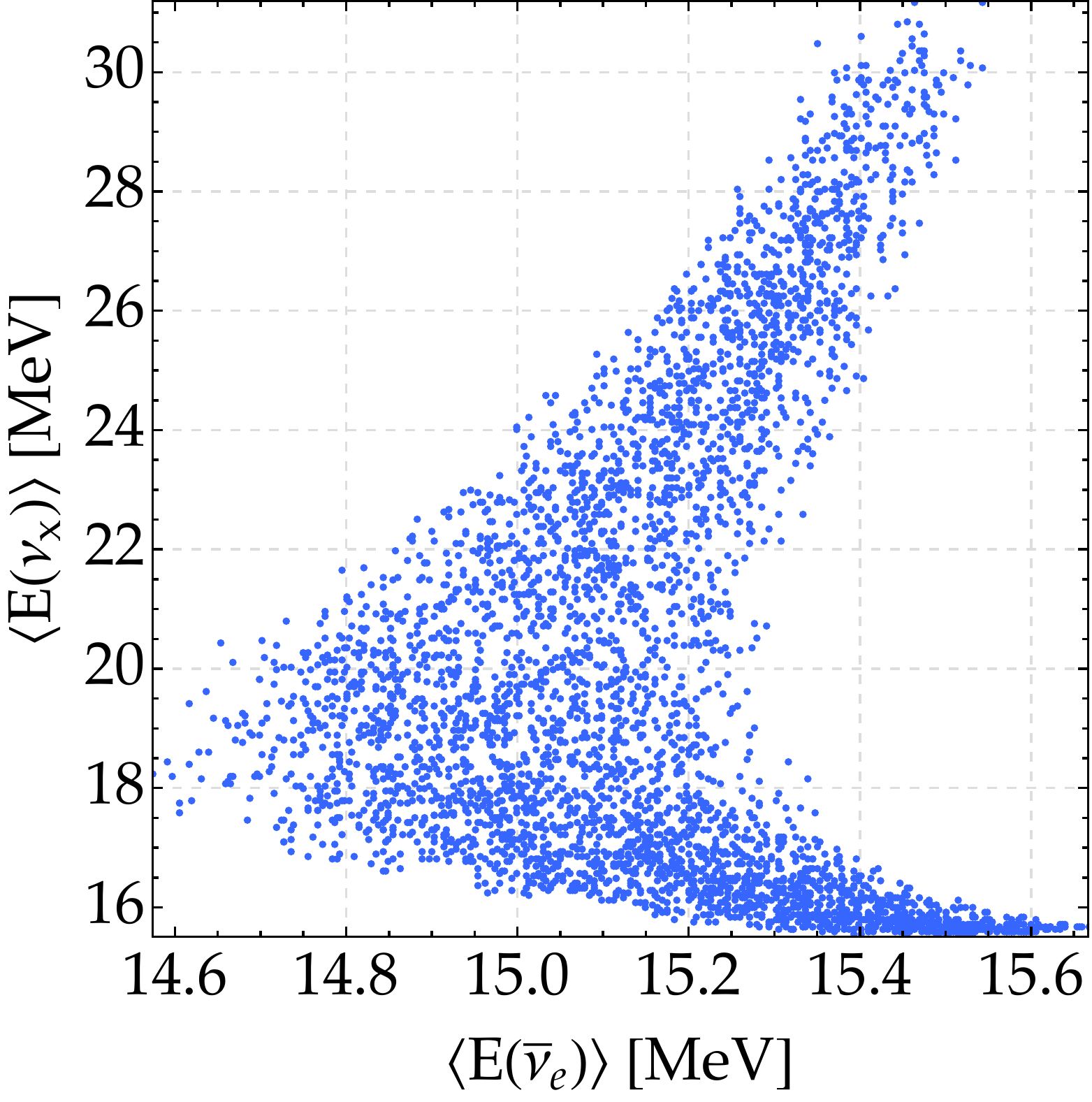}}
	\subfloat[$\langle E_{\APnue}
	\rangle$--$\mathcal{E}_{\APnue}$ projection]
	{\label{fig:mik2}
	\includegraphics[width=0.33\textwidth]{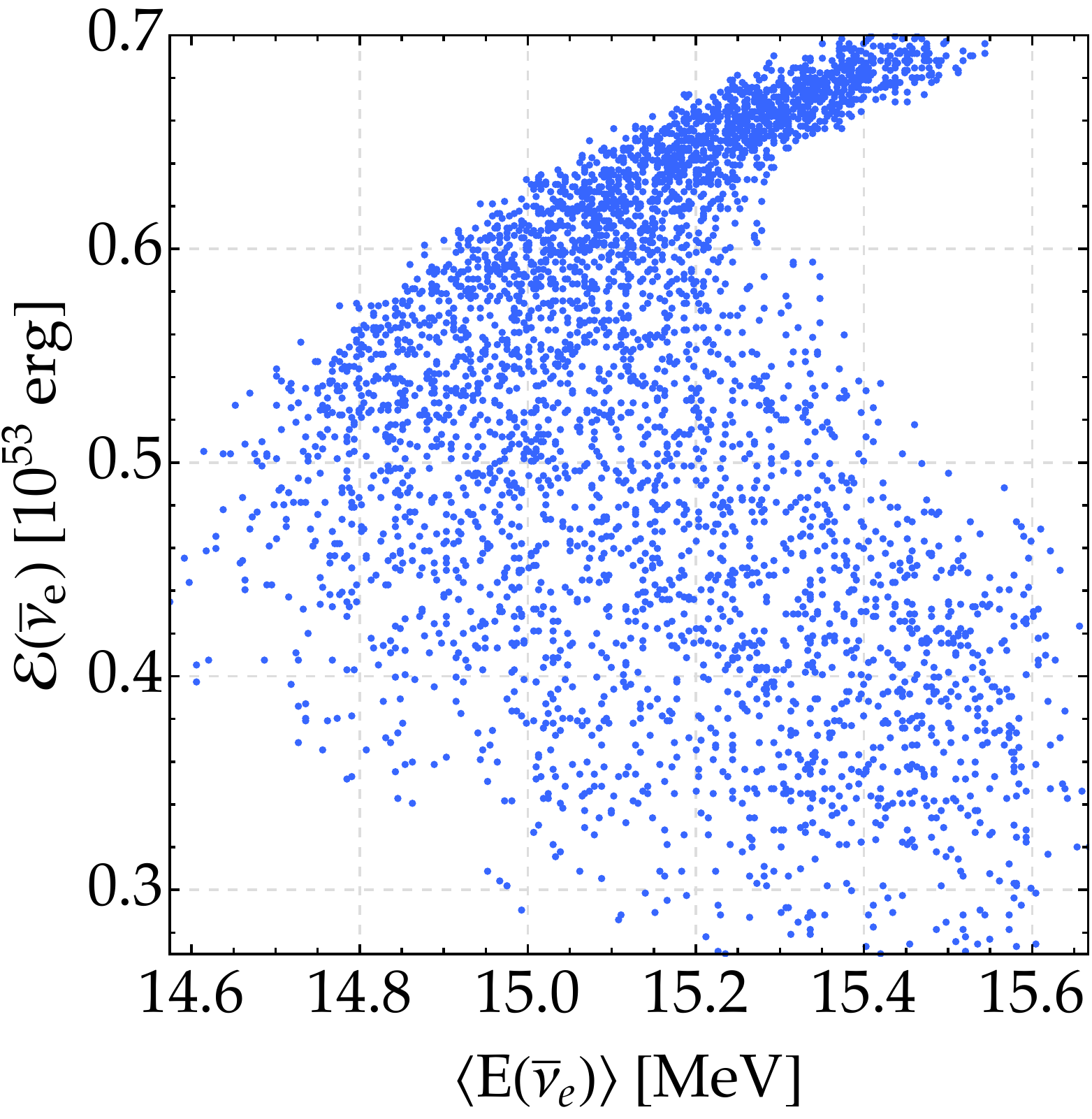}}
	\subfloat[$\langle E_{\Pnux}
	\rangle$--$\mathcal{E}_{\Pnux}$ projection]
	{\label{fig:mik3}
	\includegraphics[width=0.33\textwidth]{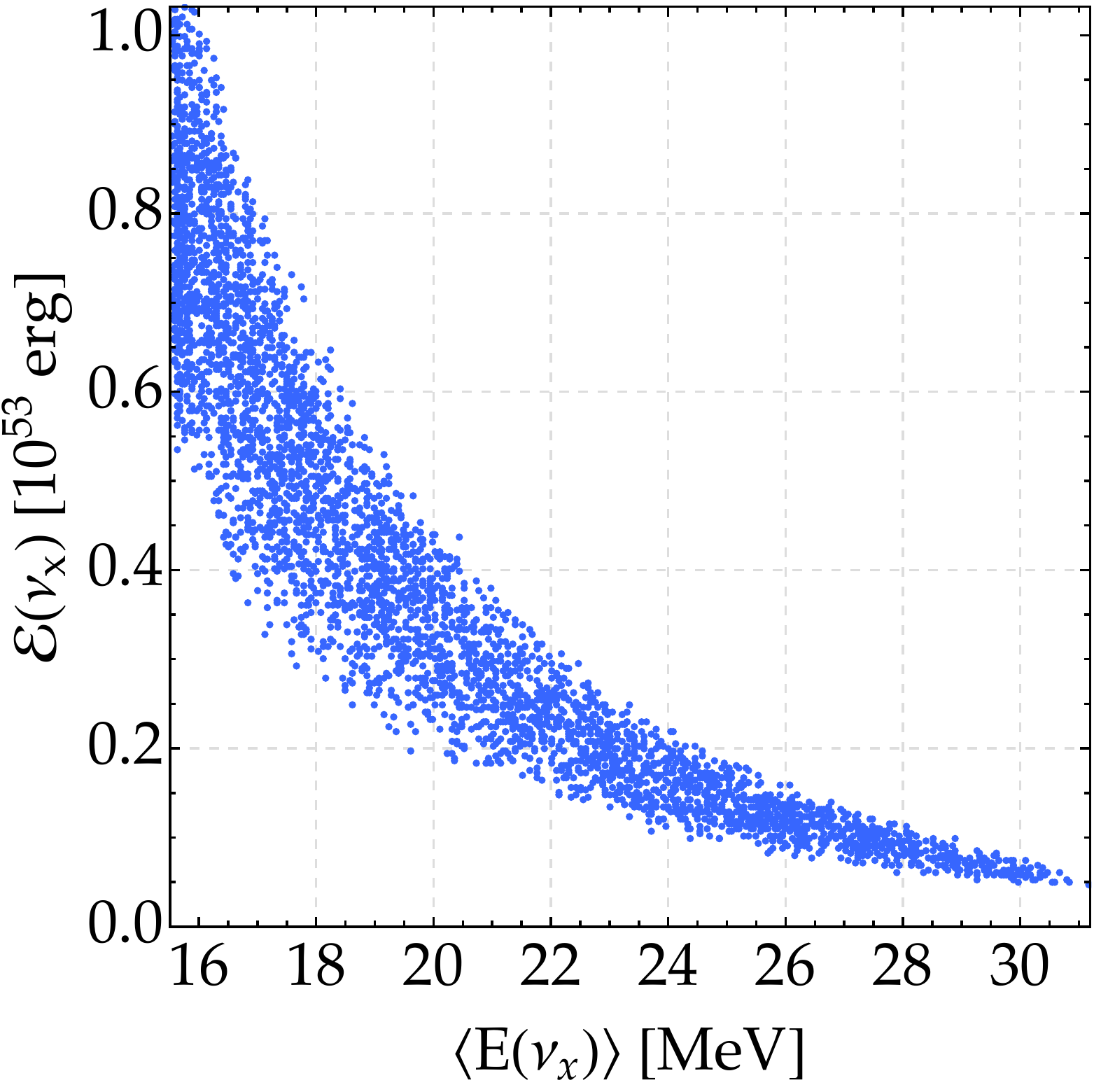}}
	\caption{Results of the 6 degrees of freedom 
	likelihood
	analysis based on IBD events in Hyper-Kamiokande.
	In order to compare our findings with those of ref.\ 
	\protect\cite{Minakata:2008nc} (figure 4)
	we make the same hypotheses but employ the Monte Carlo method
	described in the text. The results shown 
	correspond to a subset of the 6-dimensional 
	parameter space that is accepted at
	$2\sigma$ CL and projected onto: the
	$\langle E_{\APnue}\rangle$--$\langle E_{\Pnux}\rangle$
	plane \protect\subref{fig:mik1},
	$\langle E_{\APnue}\rangle$--$\mathcal{E}_{\APnue}$
	\protect\subref{fig:mik2}, and
	$\langle E_{\Pnux}\rangle$--$\mathcal{E}_{\Pnux}$
	\protect\subref{fig:mik3}.
	The prior for $\mathcal{E}_{i}$ is
	$[0.2,\, 1.2]\times 10^{53}$ erg
	while for $\langle E_i\rangle$ is
	$[8,\,32]$ MeV. }
	\label{fig:Minakata}
\end{figure}

Before performing our  main analysis including the three
detection channels, we have validated our method by reproducing
the results of ref.\ \cite{Minakata:2008nc} where a 6 degrees of
freedom statistical analysis of the IBD events in
Hyper-Kamiokande is performed.
By assuming the same fluxes and choice of the parameters as in
ref.\ \cite{Minakata:2008nc} 
we have extracted 30k 6-dimensional
points in the parameter space, accepted within 
the same confidence level, i.e.\ $2 \sigma$.
Our projected results into various
two-dimensional planes are plotted in figure~\ref{fig:Minakata}. 
As one can see by comparing figure~\ref{fig:Minakata}
with figure~4 of ref.\ \cite{Minakata:2008nc},
there is full compatibility between our results and 
those obtained by likelihood marginalization.

\section{Reconstruction of the neutrino spectra: numerical results}
\label{sec:risu}

Our analyses are performed extracting 30k points
for each of the three dataset: IBD, IBD+ES, IBD+ES+NCR.
Then, the points are projected
onto selected 2-dimensional and 1-dimensional regions, 
obtaining the distributions for the parameters defining 
the neutrino spectra,
%and information about their behavior, 
whose description and study constitute 
the aim of this section.

\subsection{Total and partial neutrino luminosities}

\begin{figure}
\centering     %%% not \center
\subfloat[Hyper-Kamiokande (standard)]{\label{fig:EtotSt}
\includegraphics[width=0.48\textwidth]{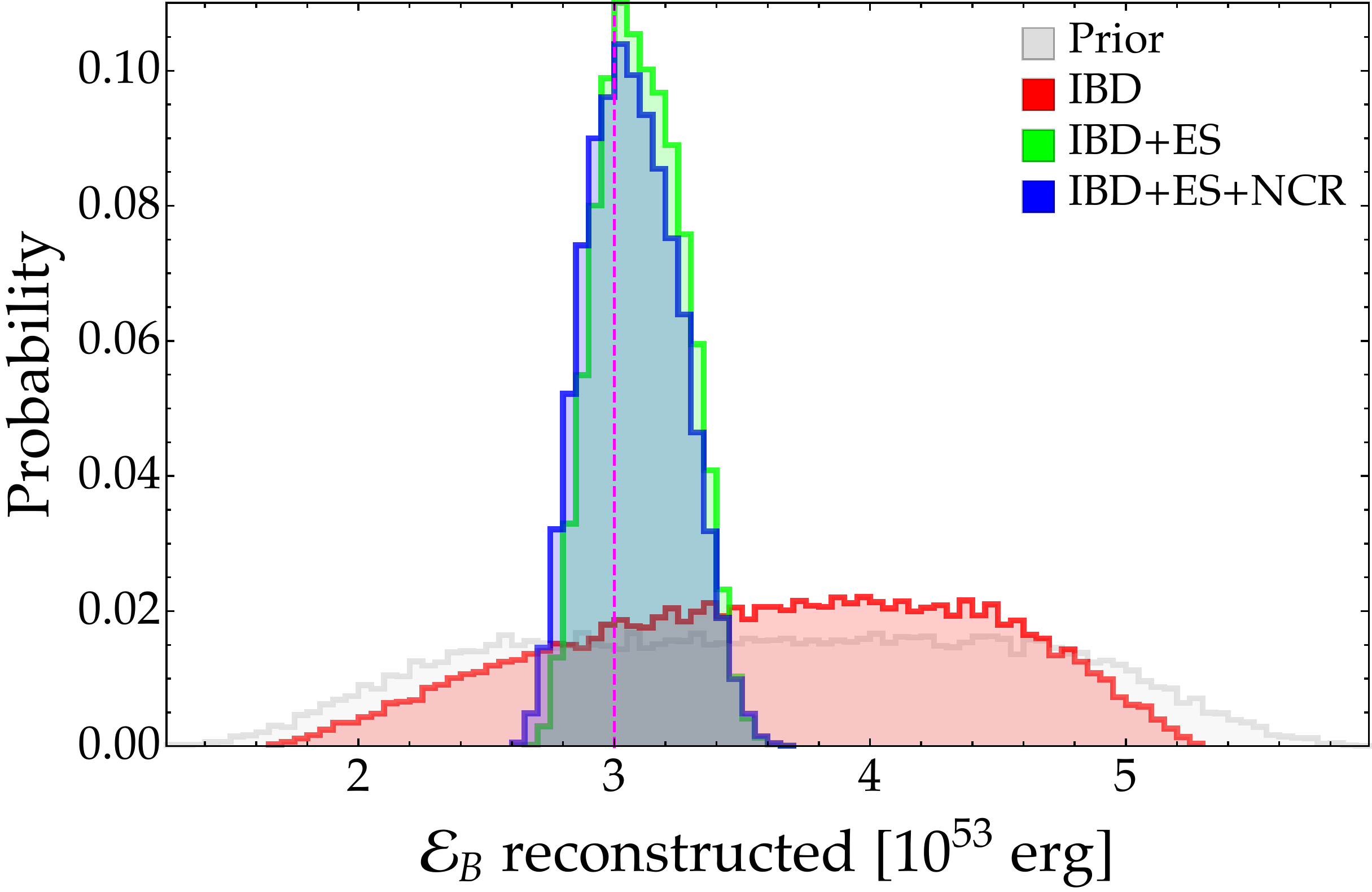}}
\subfloat[Hyper-Kamiokande (equipartition)]{\label{fig:EtotEq}\includegraphics[width=0.48\textwidth]{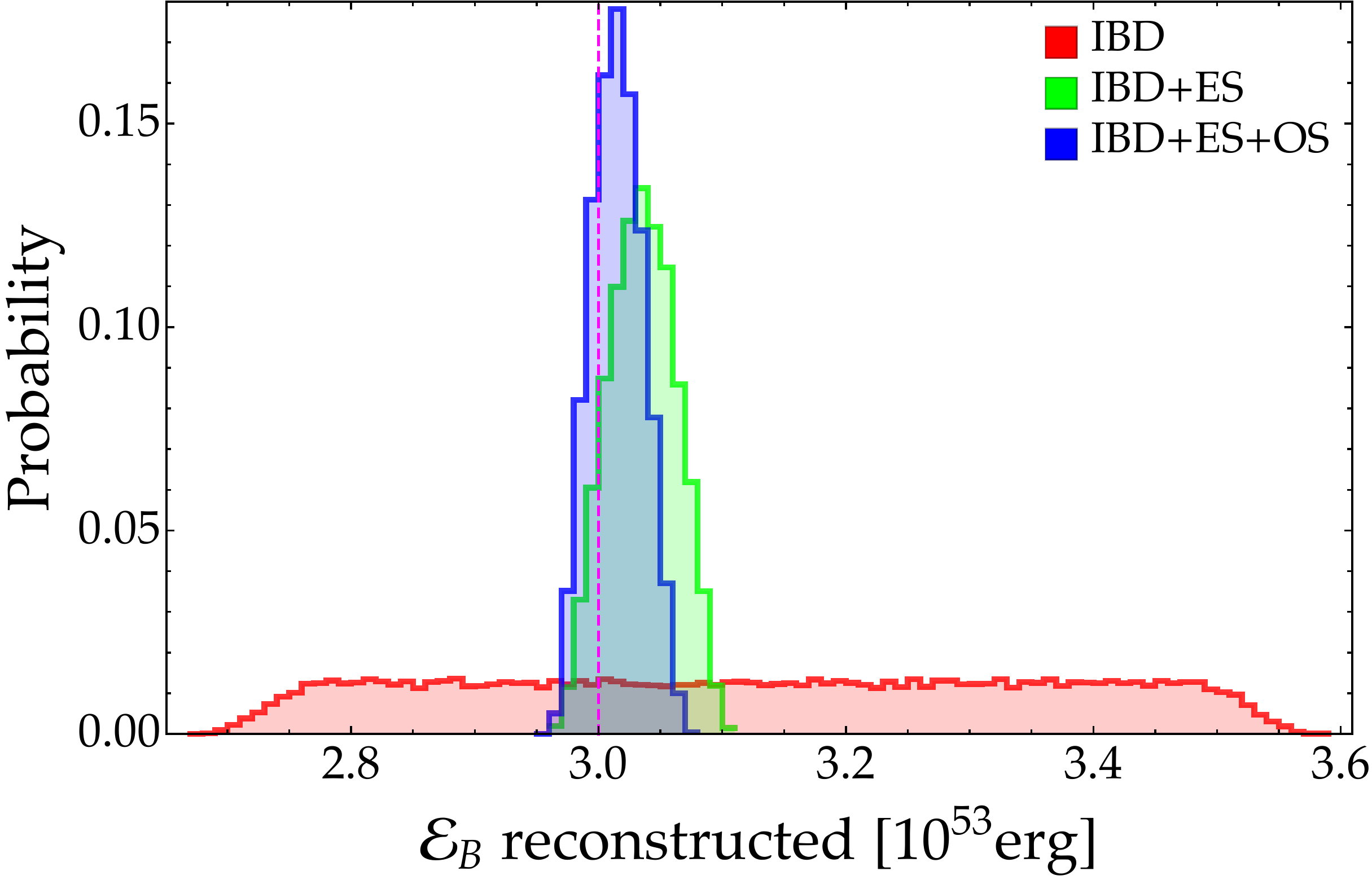}}
\caption{One-dimensional projection of the total
emitted energy $\mathcal{E}_{\mathrm{B}}$ reconstructed in
Hyper-Kamiokande from
sets of 30k extracted points at $3\sigma$ CL using
\protect\eqref{eq:EBrec}. The results correspond to the analysis either
without any constraints \protect\subref{fig:EtotSt} or
with the equipartition
{\it ansatz } \protect\subref{fig:EtotEq}.  The
different colors refer to the three analyses with increasing
information through the
number of included channels. The gray distribution
is the prior one, obtained by extracting
$\mathcal{E}_{\Pnue}$,
$\mathcal{E}_{\APnue}$,
$\mathcal{E}_{\Pnux}$ uniformly in the prior
$[0.2,1]\times\SI{e53}{\erg}$. 
The dashed magenta line
marks the true value.
Means and standard deviations of each
histogram are reported in table~\protect\ref{tab:risu}.}
\label{fig:Etot}
\end{figure}

Given an extracted point $P$, the total emitted
energy can be reconstructed as
\begin{equation}
	\label{eq:EBrec}
	\mathcal{E}_{\mathrm{B},P} =
	\mathcal{E}_{\Pnue,P} +
	\mathcal{E}_{\APnue,P}
	+ 4 \mathcal{E}_{\Pnux,P}.
\end{equation}
As can be seen from figure \ref{fig:Etot} the energy emitted
in electron neutrinos
cannot be determined from the IBD-only analysis and
remains randomly distributed inside
the prior. From the latter we extract its value in order
to determine $\mathcal{E}_{\mathrm{B}}$.
The corresponding values for each extracted
point can be gathered in a histogram. 
Figure~\ref{fig:Etot} shows the results for Hyper-Kamiokande,
without \subref{fig:EtotSt} and with \subref{fig:EtotEq}
the equipartition {\it ansatz}.
Note that the ones for Super-Kamiokande
have been published elsewhere (see figure~1 of
ref.\ \cite{GalloRosso:2017hbp}).
One can see that the IBD-only analysis leads 
to a distribution for the binding energy 
that is almost identical to the one 
that follows simply by the choice of the priors.
This result is
in agreement with the conclusions drawn by Minakata
et al.~\cite{Minakata:2008nc}. 
It is the inclusion of the other event samples
(ES and NCR) that breaks the parameter degeneracy and allows us to 
determine the total emitted energy.\footnote{Note that, in
Hyper-Kamiokande, the reconstruction is better from the IBD+ES
analysis than from the three-channels one. This proves that the
inclusion of the oxygen events does not really improve the result
but can also worsen it. In this case, the peculiar feature
might be due to the likelihoods combinations when one channel
(ES) has more events than expected and the other one (NCR) less
(see table~\ref{tab:estratti}).
}
This important conclusion differs qualitatively from the
pessimistic one drawn in ref.\ \cite{Minakata:2008nc}:
it is possible to determine the total emitted energy, 
as first shown in ref.\ \cite{GalloRosso:2017hbp} 
for the case of Super-Kamiokande.
In particular the accuracy of $\sim11\%$ with Super-Kamiokande
can be improved  to $\sim6\%$ in Hyper-Kamiokande.

\begin{sidewaystable}[p]
	\centering
	\resizebox{0.9\textwidth}{!}{%
	\begin{tabular}{|RR|Ar:Ar:Ar|Ar:Ar:Ar|c|}
		%\hline
		\cline{3-20}
		\multicolumn{2}{c|}{}
		& \multicolumn{9}{c|}{\Tstrut SUPER-KAMIOKANDE \Bstrut}
		& \multicolumn{9}{c|}{\Tstrut HYPER-KAMIOKANDE \Bstrut}\\
		\cline{1-20}
		\multicolumn{1}{|l}{\multirow{2}{*}{Param.}}
		& \multirow{ 2}{*}{$\ast$}
		& \multicolumn{3}{c}{\Tstrut\Bstrut IBD}
		& \multicolumn{3}{c}{IBD+ES}
		& \multicolumn{3}{c|}{IBD+ES+NCR}
		& \multicolumn{3}{c}{IBD}
		& \multicolumn{3}{c}{IBD+ES}
		& \multicolumn{3}{c|}{IBD+ES+NCR}\\
		\multicolumn{1}{|c}{\Bstrut} & \multicolumn{1}{c|}{}
		& \multicolumn{2}{c}{Val.}
		& \multicolumn{1}{r}{\%}
		& \multicolumn{2}{c}{Val.}
		& \multicolumn{1}{r}{\%}
		& \multicolumn{2}{c}{Val.}
		& \multicolumn{1}{r|}{\%}
		& \multicolumn{2}{c}{Val.}
		& \multicolumn{1}{r}{\%}
		& \multicolumn{2}{c}{Val.}
		& \multicolumn{1}{r}{\%}
		& \multicolumn{2}{c}{Val.}
		& \multicolumn{1}{r|}{\%} \\
		\hline
		\Tstrut \mathcal{E}_{\mathrm{B}}\Bstrut [\SI{e53}{\erg}] & 3
		& 3.40 & 0.86 & 25 & 3.27 & 0.37 & 11 & 3.18 & 0.35 & 11
		& 3.64 & 0.79 & 22 & 3.11 & 0.16 & 5.3 & 3.07 & 0.18 & 5.8
		&\multirow{11}{*}{\rotatebox[origin=c]{270}{
		\parbox[c]{1cm}{\centering STANDARD}}}\\
		\cdashline{1-20}
		\Tstrut \mathcal{E}_{\Pnue}\Bstrut [\SI{e53}{\erg}] & 0.5
		&\multicolumn{3}{c:}{---}
		& 0.60 & 0.23 & 39 & 0.57 & 0.23 & 41
		&\multicolumn{3}{c:}{---}
		& 0.57 & 0.23 & 40 & 0.56 & 0.23 & 41
		& \\
		\Tstrut \mathcal{E}_{\APnue}\Bstrut [\SI{e53}{\erg}] & 0.5
		& 0.48 & 0.11 & 22 & 0.493 & 0.051 & 10 & 0.490 & 0.050 & 10
		& 0.442 & 0.094 & 21 & 0.504 & 0.017 & 3.3 & 0.503 & 0.018 & 3.6
		& \\
		\Tstrut \mathcal{E}_{\Pnux}\Bstrut [\SI{e53}{\erg}] & 0.5
		& 0.58 & 0.23 & 40 & 0.55 & 0.10 & 18 & 0.53 & 0.09 & 18
		& 0.65 & 0.21 & 33 & 0.508 & 0.035 & 6.9 & 0.503 & 0.037 & 7.4
		& \\
		\cdashline{1-20}
		\Tstrut \langle E_{\Pnue}\rangle\Bstrut
		[\si{\mega\electronvolt}] & 9.5
		&\multicolumn{3}{c:}{---}
		& 17.6 & 6.9 & 39 & 14.0 & 5.3 & 38
		&\multicolumn{3}{c:}{---}
		& 12.4 & 4.5 & 36 & 11.0 & 4.0 & 33 & \\
		\Tstrut \langle E_{\APnue}\rangle\Bstrut
		[\si{\mega\electronvolt}] & 12
		& 12.9 & 1.6 & 12 & 11.86 & 0.82 & 6.9 & 11.99 & 0.73 & 6.1
		& 12.8 & 1.3 & 10 & 11.94 & 0.26 & 2.2 & 11.98 & 0.27 & 2.3
		& \\
		\Tstrut \langle E_{\Pnux}\rangle\Bstrut
		[\si{\mega\electronvolt}] & 15.6
		& 12.1 & 2.9 & 24 & 14.4 & 1.7 & 12 & 14.6 & 1.6 & 11
		& 13.5 & 2.0 & 15 & 15.4 & 1.1 & 7.0 & 15.7 & 1.1 & 6.9
		& \\
		\cdashline{1-20}
		\Tstrut \alpha_{\Pnue} \Bstrut & 2.5
		&\multicolumn{3}{c:}{---}
		& 2.52 & 0.58 & 23 & 2.56 & 0.57 & 22
		&\multicolumn{3}{c:}{---}
		& 2.54 & 0.57 & 23 & 2.56 & 0.57 & 22
		& \\
		\alpha_{\APnue}\Bstrut & 2.5
		& 2.36 & 0.51 & 22 & 2.13 & 0.41 & 19 & 2.22 & 0.44 & 20
		& 2.39 & 0.40 & 17 & 2.45 & 0.15 & 6.0 & 2.53 & 0.17 & 6.9
		& \\
		\Tstrut \alpha_{\Pnux}\Bstrut & 2.5
		& 2.47 & 0.57 & 23 & 2.53 & 0.56 & 22 & 2.59 & 0.55 & 21
		& 2.63 & 0.50 & 19 & 2.44 & 0.42 & 17 & 2.57 & 0.44 & 17
		& \\
		\cdashline{1-20}
		\Tstrut \kappa\Bstrut & 1
		&\multicolumn{3}{c:}{---} &\multicolumn{3}{c:}{---}
		& 0.99 & 0.11 & 11
		&\multicolumn{3}{c:}{---} &\multicolumn{3}{c:}{---}
		& 0.90 & 0.06 & 6.3 & \\
		\hline
 		%%%%%%%%%%%%%%%%%%%%%%%%%%%%%%%%%%%%%%%%%%%%%%%%%%%%%%%%
 		% EQUIPARTIZIONE
 		%%%%%%%%%%%%%%%%%%%%%%%%%%%%%%%%%%%%%%%%%%%%%%%%%%%%%%%%
 		\Tstrut \mathcal{E}_{\mathrm{B}}\Bstrut [\SI{e53}{\erg}] & 3
		& 3.15 & 0.25 & 7.9 & 3.06 & 0.10 & 3.4 & 3.023 & 0.095 & 3.1
		&3.13 & 0.23 & 7.4 & 3.035 & 0.027 & 0.89 & 3.015 & 0.021 & 0.68
		&\multirow{11}{*}{\rotatebox[origin=c]{270}{
		\parbox[c]{2cm}{\centering EQUIPARTITION}}}\\
		\cdashline{1-20}
		\Tstrut \mathcal{E}_{\Pnue}\Bstrut [\SI{e53}{\erg}] & 0.5
		&\multicolumn{3}{c:}{---}
		& 0.511 & 0.017 & 3.4 & 0.504 & 0.016 & 3.1
		&\multicolumn{3}{c:}{---}
		& 0.5059 & 0.0045 & 0.89 & 0.5025 & 0.0034 & 0.68
		&\\
		\Tstrut \mathcal{E}_{\APnue}\Bstrut [\SI{e53}{\erg}] & 0.5
		& 0.511 & 0.019 & 3.7
		& 0.511 & 0.017 & 3.4 & 0.504 & 0.016 & 3.1
		& 0.5057 & 0.0047 & 0.92
		& 0.5059 & 0.0045 & 0.89 & 0.5025 & 0.0034 & 0.68
		& \\
		\Tstrut \mathcal{E}_{\Pnux}\Bstrut [\SI{e53}{\erg}] & 0.5
		& 0.511 & 0.019 & 3.7
		& 0.511 & 0.017 & 3.4 & 0.504 & 0.016 & 3.1
		& 0.5057 & 0.0047 & 0.92
		& 0.5059 & 0.0045 & 0.89 & 0.5025 & 0.0034 & 0.68
		& \\
		\cdashline{1-20}
		\Tstrut \langle E_{\Pnue}\rangle\Bstrut
		[\si{\mega\electronvolt}] & 9.5
		&\multicolumn{3}{c:}{---}
		& 17.9 & 6.9 & 39 & 14.3 & 5.2 & 37
		&\multicolumn{3}{c:}{---}
		& 12.8 & 4.4 & 34 & 10.6 & 3.0 & 28
		& \\
		\Tstrut \langle E_{\APnue}\rangle\Bstrut
		[\si{\mega\electronvolt}] & 12
		& 13.0 & 1.4 & 11 & 11.72 & 0.68 & 5.8 & 11.92 & 0.65 & 5.4
		& 12.59 & 0.81 & 6.5 & 11.92 & 0.26 & 2.2 & 11.94 & 0.21 & 1.7
		& \\
		\Tstrut \langle E_{\Pnux}\rangle\Bstrut
		[\si{\mega\electronvolt}] & 15.6
		& 12.1 & 2.8 & 23 & 14.8 & 1.6 & 11 & 14.8 & 1.5 & 10
		& 13.7 & 2.1 & 16 & 15.4 & 1.0 & 6.5 & 15.84 & 0.80 & 5.0
		&\\
		\cdashline{1-20}
		\Tstrut \alpha_{\Pnue} \Bstrut & 2.5
		&\multicolumn{3}{c:}{---}
		& 2.52 & 0.58 & 23 & 2.55 & 0.57 & 22
		&\multicolumn{3}{c:}{---}
		& 2.54 & 0.57 & 23 & 2.57 & 0.57 & 22
		&\\
		\alpha_{\APnue}\Bstrut & 2.5
		& 2.35 & 0.52 & 22 & 2.09 & 0.39 & 18 & 2.20 & 0.43 & 20
		& 2.25 & 0.32 & 14 & 2.44 & 0.16 & 6.4 & 2.53 & 0.13 & 5.3
		&\\
		\Tstrut \alpha_{\Pnux}\Bstrut & 2.5
		& 2.48 & 0.57 & 23 & 2.61 & 0.55 & 21 & 2.63 & 0.55 & 21
		& 2.80 & 0.51 & 18 & 2.46 & 0.40 & 16 & 2.61 & 0.34 & 13
		&\\
		\cdashline{1-20}
		\Tstrut \kappa\Bstrut & 1
		&\multicolumn{3}{c:}{---} &\multicolumn{3}{c:}{---}
		& 0.99 & 0.11 & 11
		&\multicolumn{3}{c:}{---} &\multicolumn{3}{c:}{---}
		& 0.953 & 0.098 & 10
		&\\
 		\hline
 	\end{tabular}
	}
	\caption{Reconstruction of the emission parameters in
	Super-Kamiokande (left block) and Hyper-Kamiokande
	(right block), without (top rows)
	and with (bottom rows) equipartition {\it ansatz}.
	%The supernova neutrino fluences are parametrized by pinched
	%Fermi-Dirac distributions that have been modified by the MSW
	%mechanism in the case of normal mass ordering.
	%Further
	Considerations on the emission parameters can be
	found in section~\protect\ref{sec:1:2}.
	The three analyses differ for the
	number of detection channels included: IBD, IBD+ES, IBD+ES+NCR
	(see section~\protect\ref{sec:1:3}).
	The $\kappa$ parameter takes into
	account the theoretical uncertainty on the neutral current
	neutrino-oxygen cross section (see section~\ref{sec:1:2}).
	In each cell we report mean, standard deviation and percentage
	accuracy obtained from 30k extracted random points accepted
	at $3\sigma$ CL.
	%from their distributions. 
	The second column marked by the asterisk reports the true
	values of the emission parameters used in the analysis.
	}
 	\label{tab:risu}
\end{sidewaystable}

If an exact equipartition
held true, Super-Kamiokande would be 
able to reach a precision of few percent while Hyper-Kamiokande
could  reconstruct the total energy with an accuracy less than 
$1\%$ --- namely $0.7\%$.\footnote{As can be seen in
figure~\ref{fig:EtotEq} and table~\ref{tab:risu},
the total emitted energy is
slightly over-estimated, remaining still fully compatible within the
statistical error. This behavior can be due to the random
extraction of more ES events than expected
(see table~\ref{tab:estratti}).}
Let us repeat that the hypothesis of
an exact equipartition nowadays should be regarded as a very
aggressive  assumption, since current simulations indicate
that it should hold true only within a factor of 2
(see e.g.\ refs.\ \cite{Keil:2002in,Keil:2003sw,Raffelt:2005fb}).

\begin{figure}[p]
\centering     %%% not \center
\subfloat[Super-Kamiokande]{\label{fig:skter}
\includegraphics[width=0.48\textwidth]{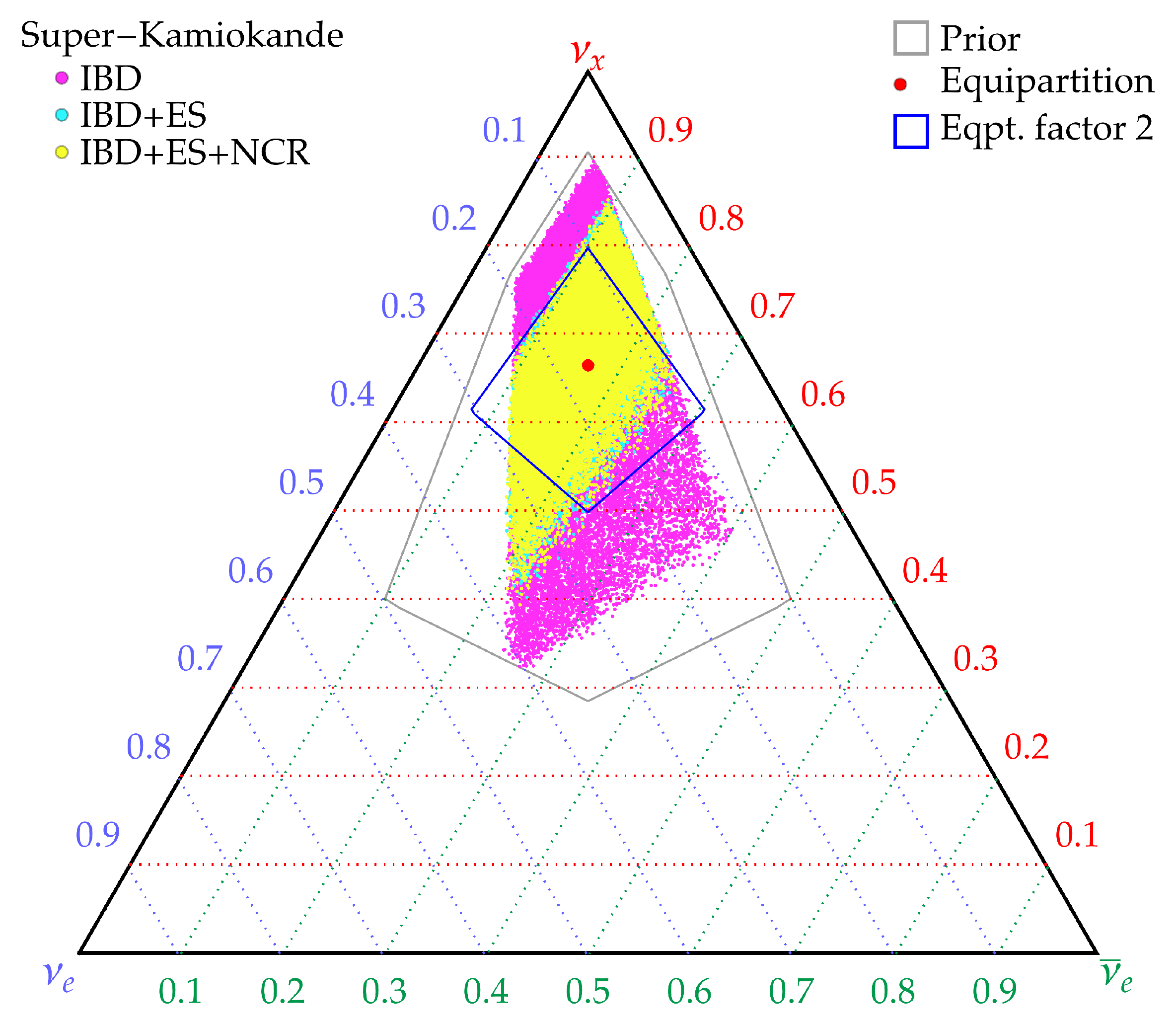}}
\subfloat[Hyper-Kamiokande]{\label{fig:hkter}
\includegraphics[width=0.48\textwidth]{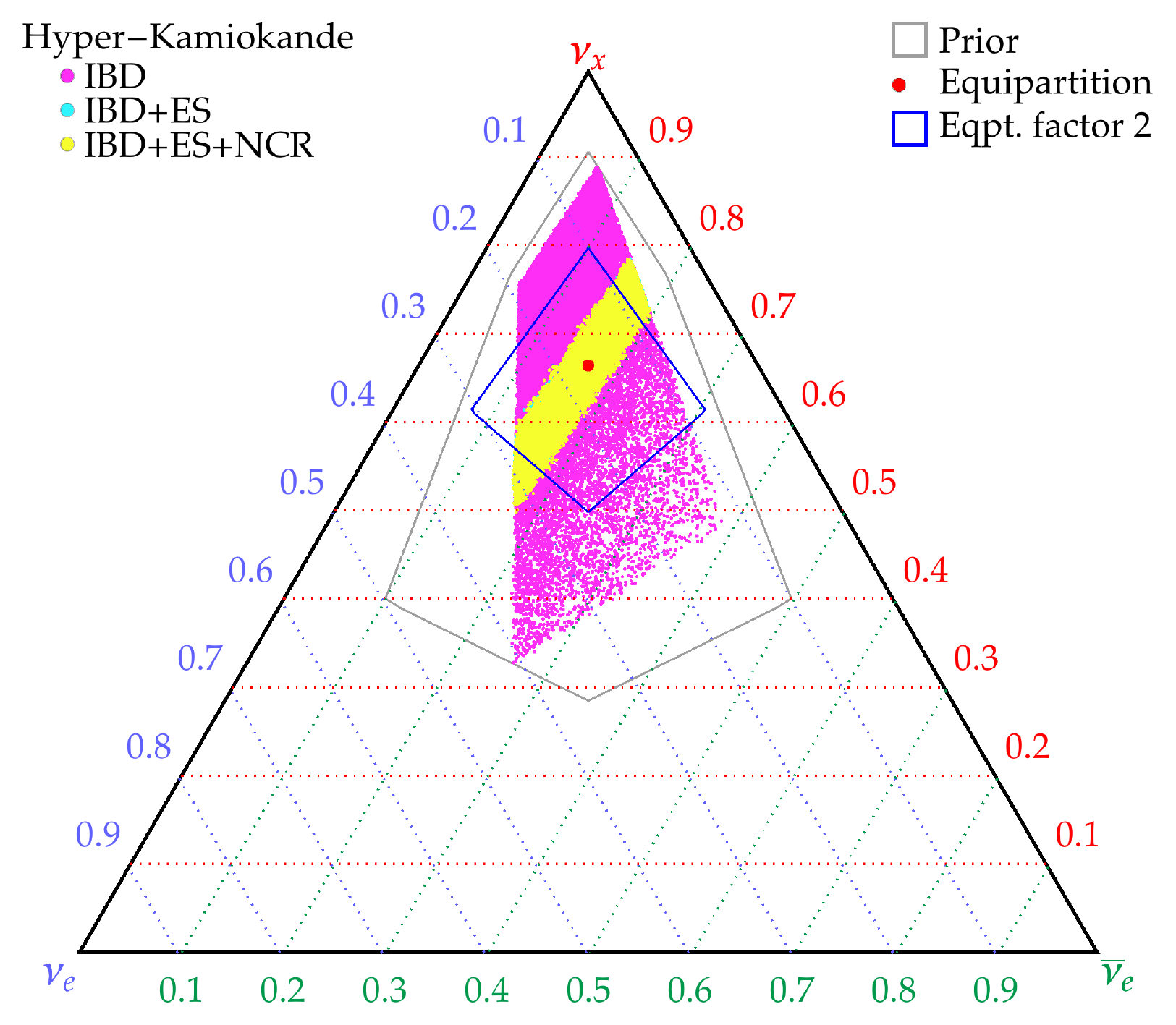}}
\caption{Ternary plot for Super-Kamiokande
\protect\subref{fig:skter} and Hyper-Kamiokande
\protect\subref{fig:hkter} where the three axes correspond to
the amount of $\mathcal{E}_{\Pnue}$,
$\mathcal{E}_{\APnue}$ and $4\mathcal{E}_{\Pnux}$ for
each extracted point. Different colors represent
the three analyses performed with increasing
information gained through the addition 
of detection channels. The gray line
marks the allowed \textit{a priori} region, obtained extracting
$\mathcal{E}_{\Pnue}$,
$\mathcal{E}_{\APnue}$,
$\mathcal{E}_{\Pnux}$ uniformly in the prior
$[0.2,1]\times\SI{e53}{\erg}$. The blue line, on the
other hand contains the points for which the equipartition
hypothesis holds true within a factor of 2.
The red dot indicates the value corresponding to
the equipartition {\it ansatz}. Each set contains
30k extracted points at $3\sigma$ CL. 
}
\label{fig:ternario}
%\end{figure}
%\begin{figure}
\centering     %%% not \center
\subfloat[IBD+ES+NCR]{\label{fig:scate2}
\includegraphics[width=0.48\textwidth]{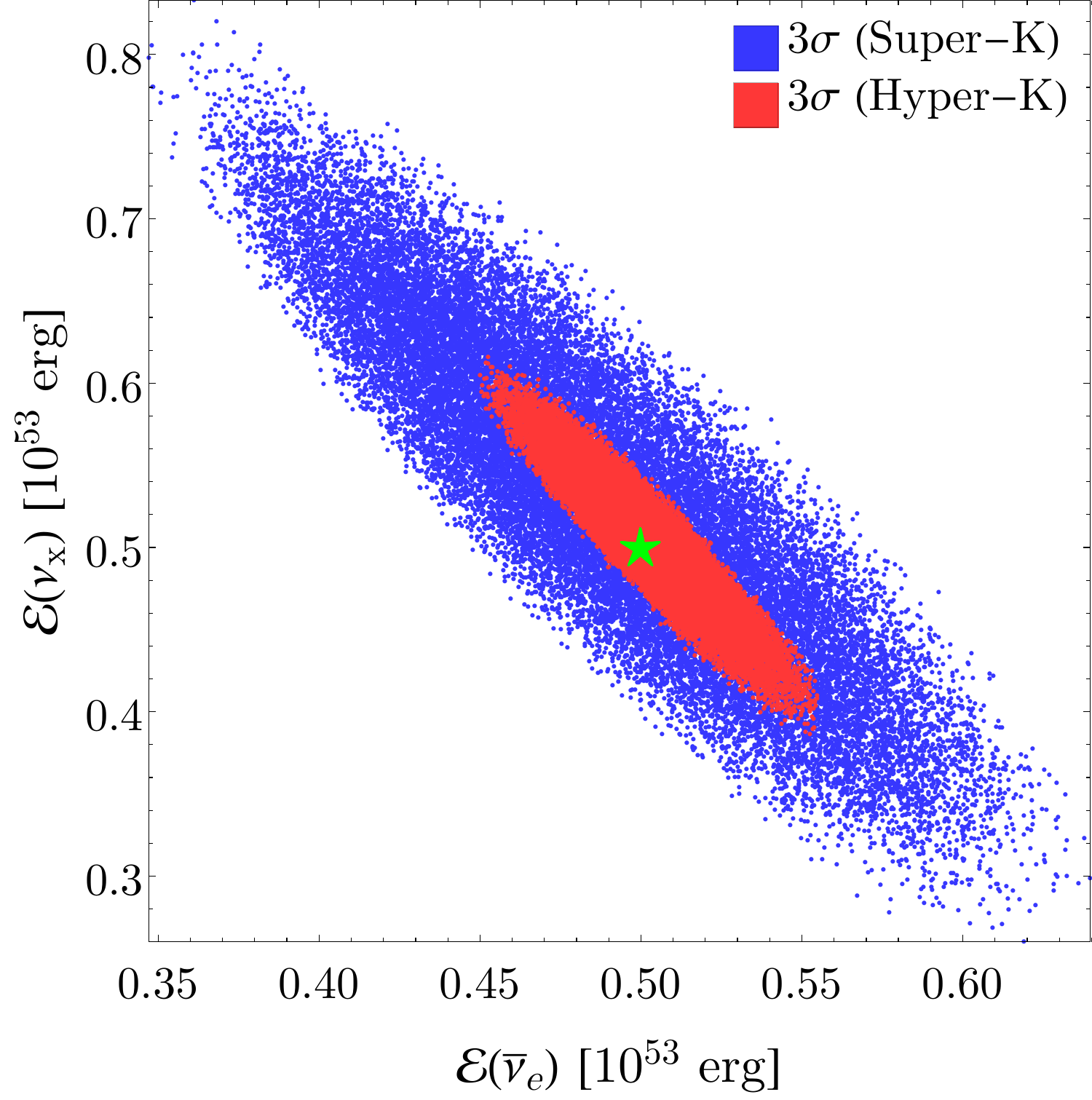}}
\subfloat[IBD+ES+NCR]{\label{fig:scate1}
\includegraphics[width=0.49\textwidth]{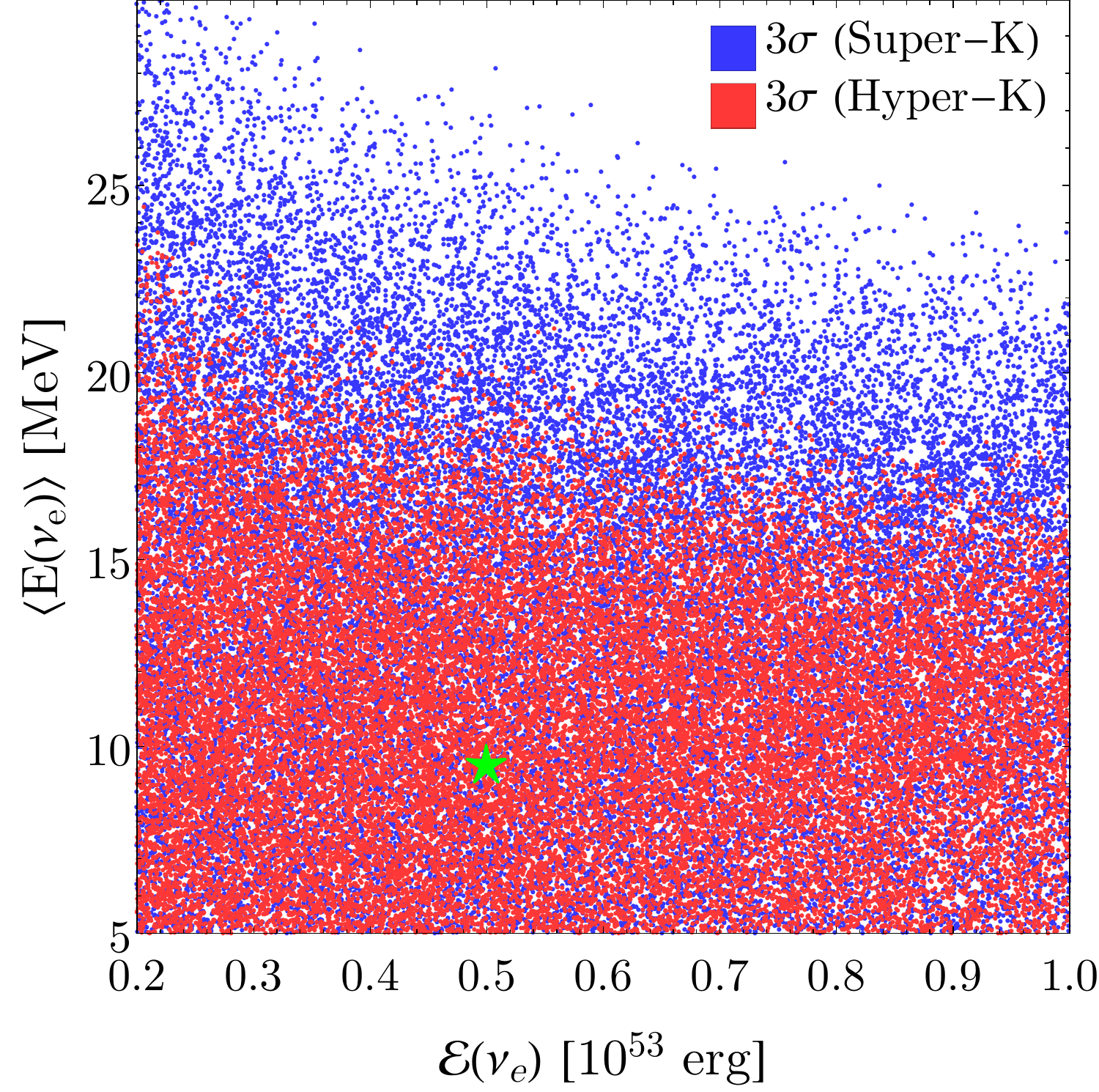}}
\caption{Projection of the extracted points onto
the $\mathcal{E}_{\APnue}$--$\mathcal{E}_{\Pnux}$ plane
\protect\subref{fig:scate2} 
and $\mathcal{E}_{\Pnue}$--$\langle E_{\Pnue}
\rangle$ \protect\subref{fig:scate1}
plane. The blue dots refer to
Super-Kamiokande while the red ones to Hyper-Kamiokande.
Each set contains 30k points, accepted at $3\sigma$ CL.
The three detection channels have been considered.
Notice that the prior for $\mathcal{E}_i$ is
$[0.2,1]\times\SI{e53}{\erg}$ and $[5,30]\,\si{\mega\electronvolt}$
for $\langle E_i \rangle$. The green star marks the true value.
}%ddd
\label{fig:scate}
\end{figure}

A way of representing the fraction of the total energy
carried out by each species is by plotting the 
reconstructed emitted energies per flavor
on a ternary diagram. This is shown in figure~\ref{fig:skter}
for Super-Kamiokande and figure \ref{fig:hkter} for
Hyper-Kamiokande. As one can see the ternary diagrams are
not much instructive since the points are distributed over the
whole region.
If we assume the factor of 2 of uncertainty is a
reliable estimate, we  can decrease the size of the region where
the point are extracted, and keep only those that fall inside the
blue line. This additional assumption, however, does not improve
significantly the results obtained with the three-channels analysis.
For instance, the three-channels analysis in Hyper-Kamiokande
gives a total emitted energy of $\left(3.12\pm 0.15\right)
\times\SI{e53}{\erg}$, that corresponds to a slightly better
accuracy ($\sim 5\%$) just because the reconstructed
value is bigger.
%In fact the accuracy remains almost the same
%(table~\ref{tab:risu}).

The projected emitted energy for $\nu_x$ as a function of
the $\overline{\nu}_e$ one is shown in figure~\ref{fig:scate}.
One can see that Hyper-Kamiokande
will be able to measure the energy emitted in  $\nu_x$ species 
(4 of them, out of 6) with an accuracy of  $\sim7\%$,
instead of the $\sim18 \%$ that Super-Kamiokande
should achieve (table~\ref{tab:risu}).
As for the $\overline{\nu}_{\mathrm{e}}$ the measurement of the
corresponding (integrated) luminosity goes from
$\sim10 \%$ in Super-Kamiokande to $\sim4 \%$ in Hyper-Kamiokande.
The inclusion of the
$\nu_{\mathrm{e}}$ and $\overline{\nu}_{\mathrm{e}}$
species to the $\nu_x$ ones leads to a 
slight decrease of the
uncertainty on the total emitted energy, that has already mentioned, 
amounts to $5\%$.

Studying the total energy distributions
for each neutrino flavor, we notice that the main source of
uncertainties is attributable to the fact 
that no specific detection channel sensitive to electron
neutrinos is added in the analysis.
This can be seen from table~\ref{tab:estratti} since the
fraction of $\nu_{\mathrm{e}}$ events is small compared to the
$\overline{\nu}_{\mathrm{e}}$ or $\nu_x$ ones.
As a consequence, the reconstruction of the $\nu_{\mathrm{e}}$
 emitted energy
{\em does not} improve from Super-Kamiokande to Hyper-Kamiokande;
the electron neutrino properties remain
largely undetermined. This is true for the electron neutrino
luminosity as well as for the mean energy and pinching parameter.

\subsection{Mean energies and pinching}
\begin{figure}[t]
\centering     %%% not \center
\subfloat[$\langle E_{\APnue} \rangle$
in Super-Kamiokande]{\label{fig:ske2}
\includegraphics[width=0.48\textwidth]{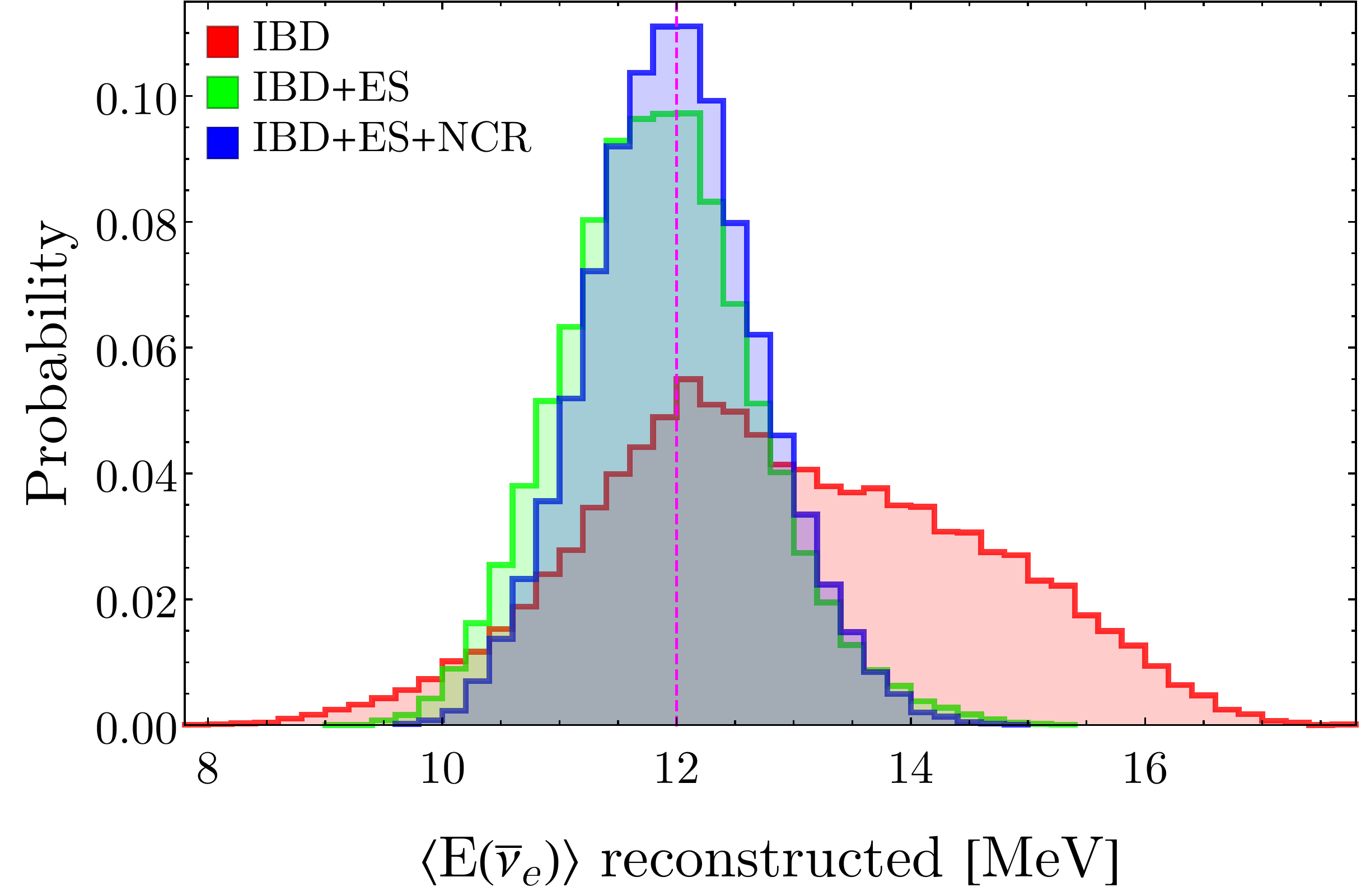}}
\subfloat[$\langle E_{\APnue} \rangle$
in Hyper-Kamiokande]{\label{fig:hke2}
\includegraphics[width=0.48\textwidth]{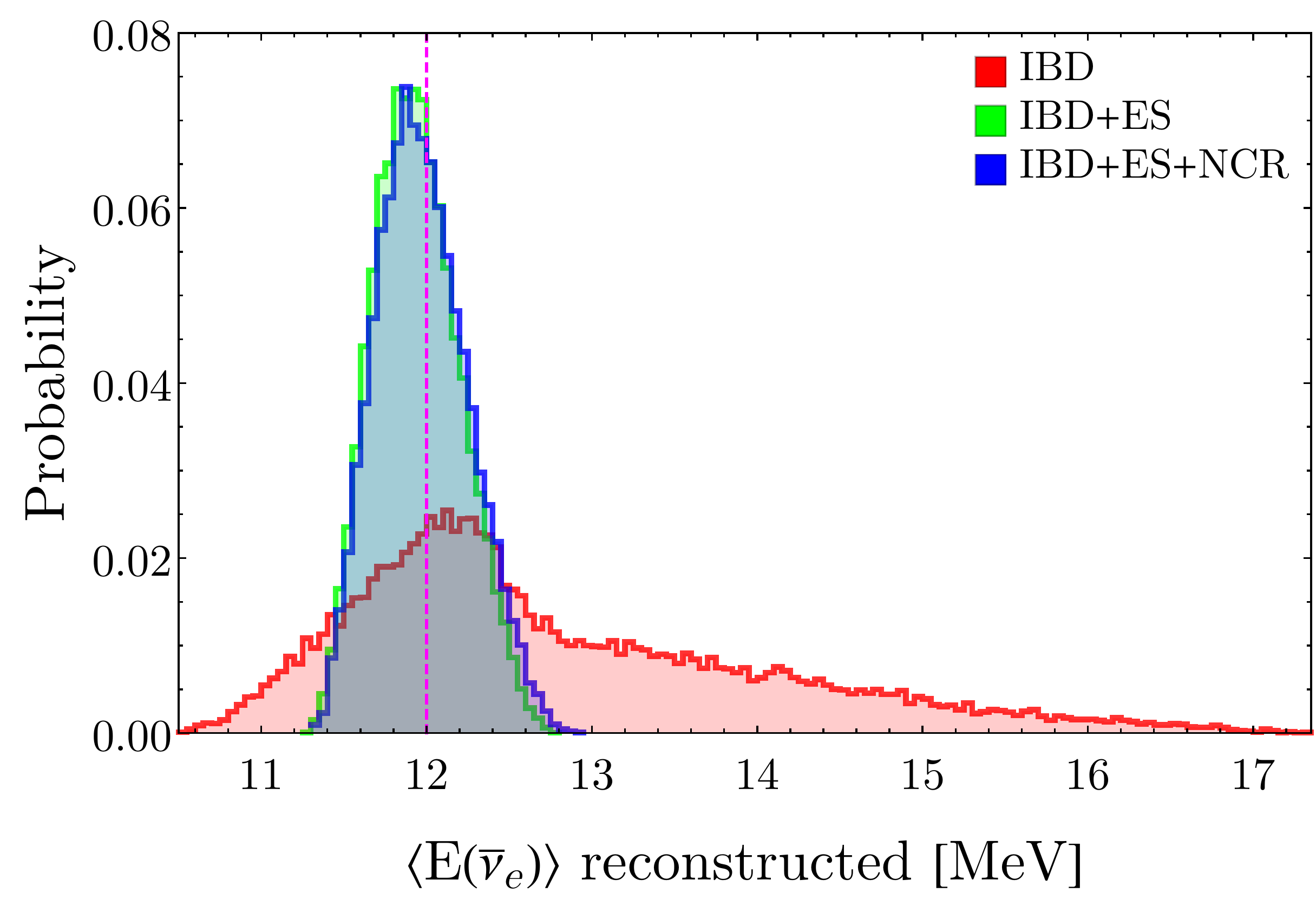}}\\
\subfloat[$\langle E_{\Pnux} \rangle$
in Super-Kamiokande]{\label{fig:ske3}
\includegraphics[width=0.48\textwidth]{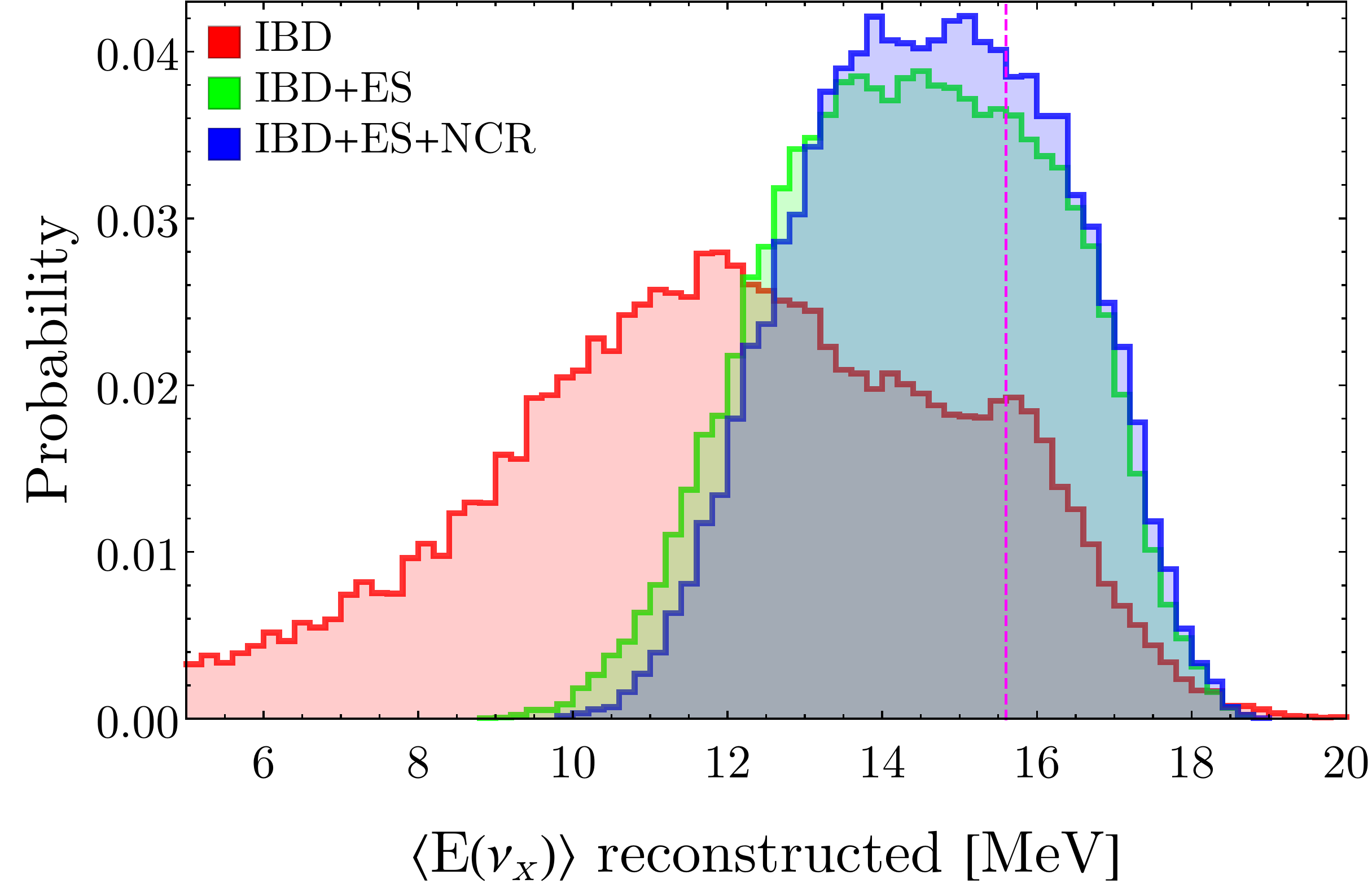}}
\subfloat[$\langle E_{\Pnux} \rangle$
in Hyper-Kamiokande]{\label{fig:hke3}
\includegraphics[width=0.48\textwidth]{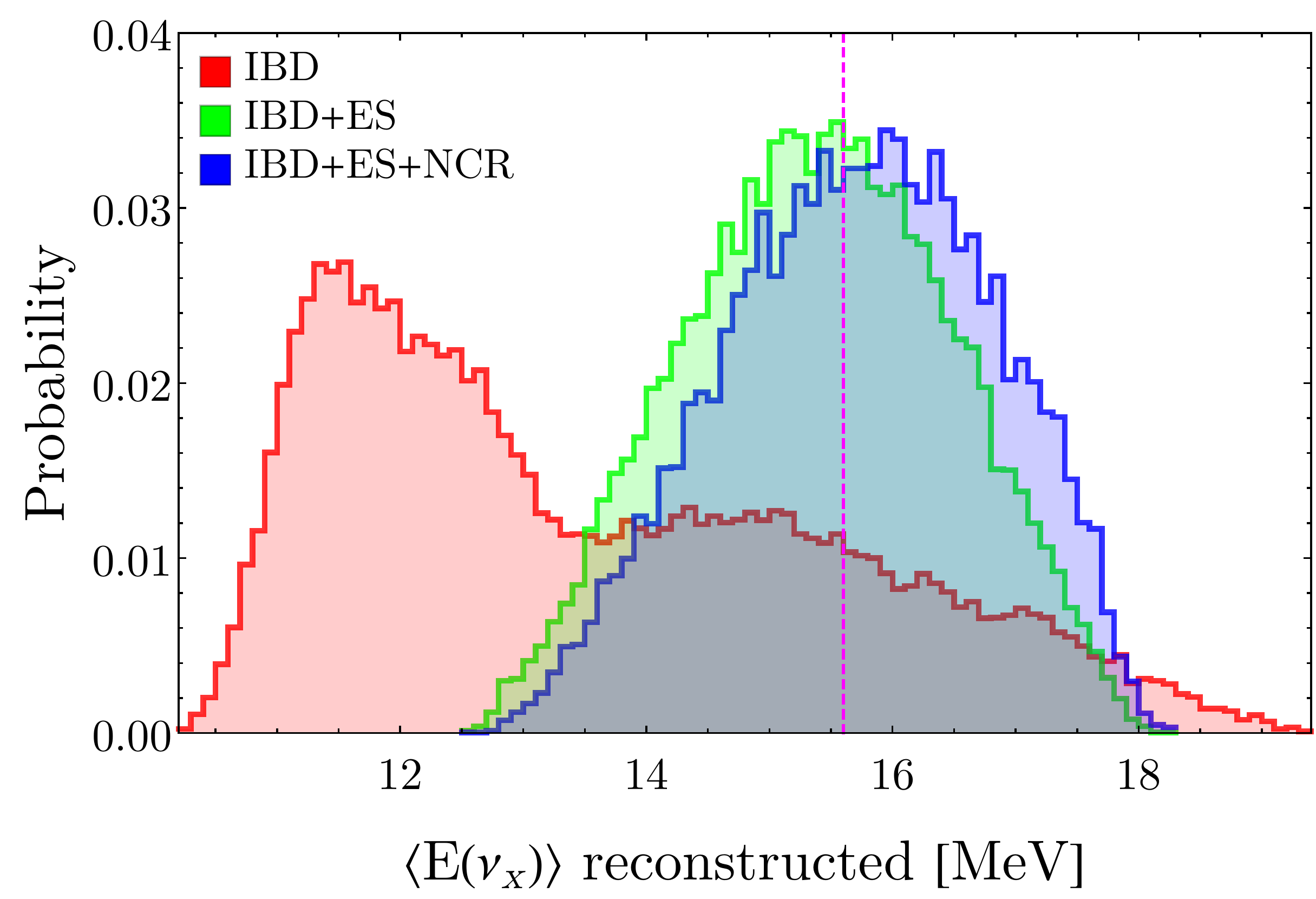}}
\caption{Distributions of reconstructed $\langle E_{\APnue}\rangle$
(first row) and $\langle E_{\Pnux}\rangle$ (second row) in
Super-Kamiokande (left) and Hyper-Kamiokande (right).
Different colors represent the amount of information used through
the number of detection channels. Each histogram contains
30k extracted points, accepted at $3\sigma$ CL.
The dashed magenta line marks
the true value. Means and standard deviations of those
histograms are reported in table~\protect\ref{tab:risu}.
}
\label{fig:mene}
\end{figure}

The mean energies of $\overline{\nu}_{\mathrm{e}}$ and
$\nu_x$ can be precisely
reconstructed. Their values are plotted in figure~\ref{fig:mene},
for the two detectors and as a function of number of detection
channels. The corresponding averages and
standard deviations are reported in table~\ref{tab:risu}.
A precision of  $\sim2 \%$ in Hyper-Kamiokande ($\sim6 \%$ in
Super-Kamiokande) and $\sim 7 \%$ in Hyper-Kamiokande ($\sim 11 \%$
in Super-Kamiokande) can be reached for
$\overline{\nu}_{\mathrm{e}}$ and $\nu_x$ respectively.
Again, this result is achieved thanks to the combination of the
main channels, i.e.\ inverse beta decay and neutrino elastic
scattering on electrons. In fact, table \ref{tab:risu} shows that
IBD-only is not sufficient to determine precisely
even the energy of the $\overline{\nu}_{\mathrm{e}}$,
despite the high statistics (see table~\ref{tab:estratti}).
The inclusion of elastic scattering interactions in the analysis 
improves strongly the outcome. As expected, the
$\langle E_{\APnue} \rangle$ improves more than $\nu_x$ due to
the higher percentage of survived
$\overline{\nu}_{\mathrm{e}}$ after the MSW effect. 

The reconstructed value  of $\langle E_{\Pnux} \rangle$
in Super-Kamiokande turns out to be slightly below the expectations. 
While this is within the statistical uncertainty,
one reason that favors this behavior is 
the flux parameters degeneracy, that allows
an interplay between the normalization of the
spectrum $\mathcal{E}_{\Pnux}$,  its first momentum 
$\langle E_{\Pnux}\rangle$ and its width
$\alpha_{\Pnux}$. For instance, in the analysis based on three
detection channels,
a slightly underestimated value  of $\langle E_{\Pnux}\rangle$
corresponds to a slight overestimation of $\alpha_{\Pnux}$ and of 
$\mathcal{E}_{\Pnux}$.

The projected results for $\nu_{\mathrm{e}}$
on the $\mathcal{E}_{\Pnue}$--$\langle E_{\Pnue}
\rangle$ plane are given in figure~\ref{fig:scate1}. As one can see
their distribution is almost uniform inside the prior.
This shows that the $\nu_{\mathrm{e}}$ average energies 
are to large extent undetermined since
$\langle E_{\Pnue} \rangle$ goes almost uniformly from
5 to \SI{25}{\mega\electronvolt} for Super-Kamiokande
and from 5 to \SI{20}{\mega\electronvolt} for Hyper-Kamiokande. 

\begin{figure}[t]
\centering     %%% not \center
\subfloat[IBD]{\label{fig:scate3}
\includegraphics[width=0.48\textwidth]{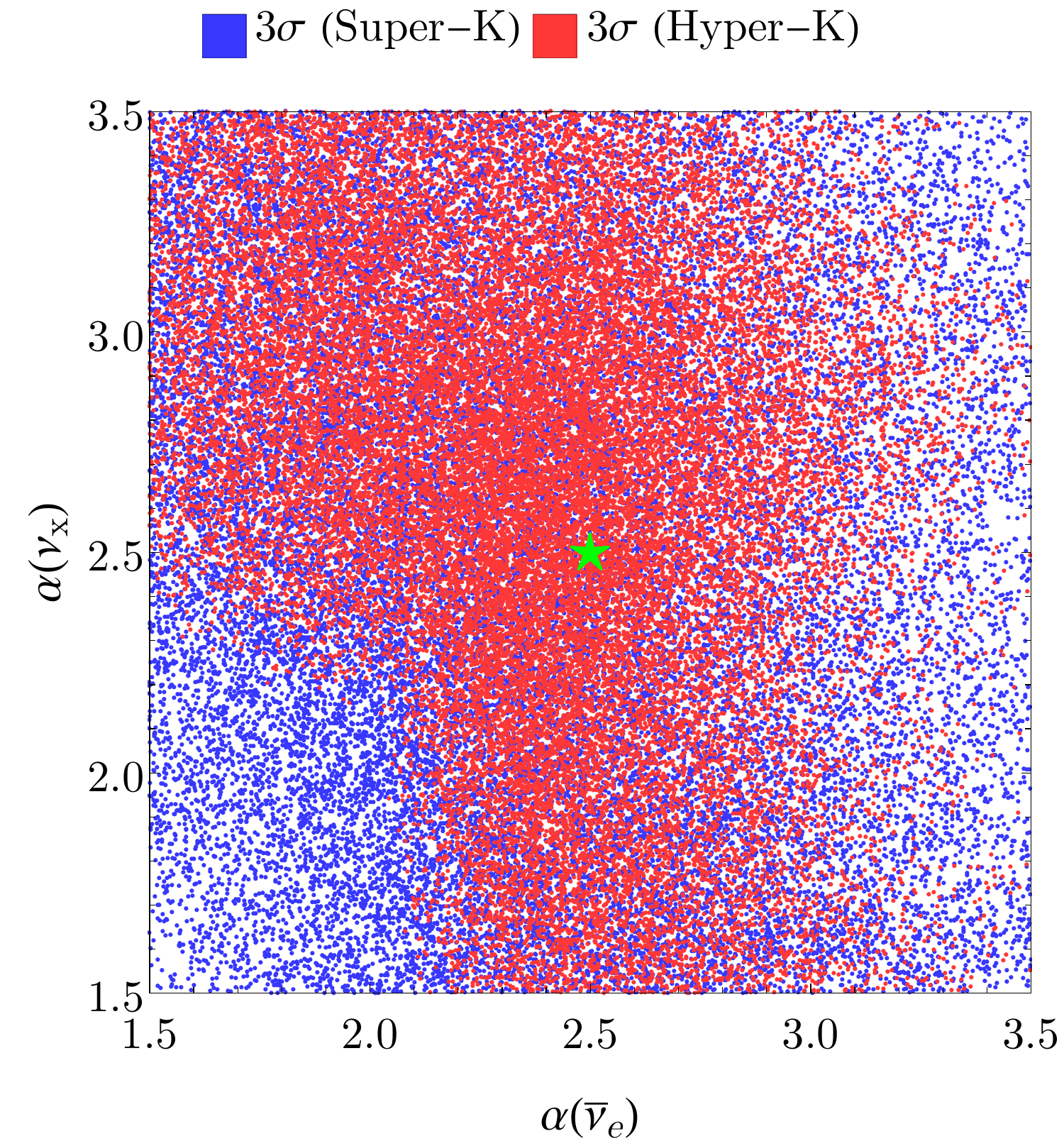}}
\subfloat[IBD+ES+NCR]{\label{fig:scate4}
\includegraphics[width=0.48\textwidth]{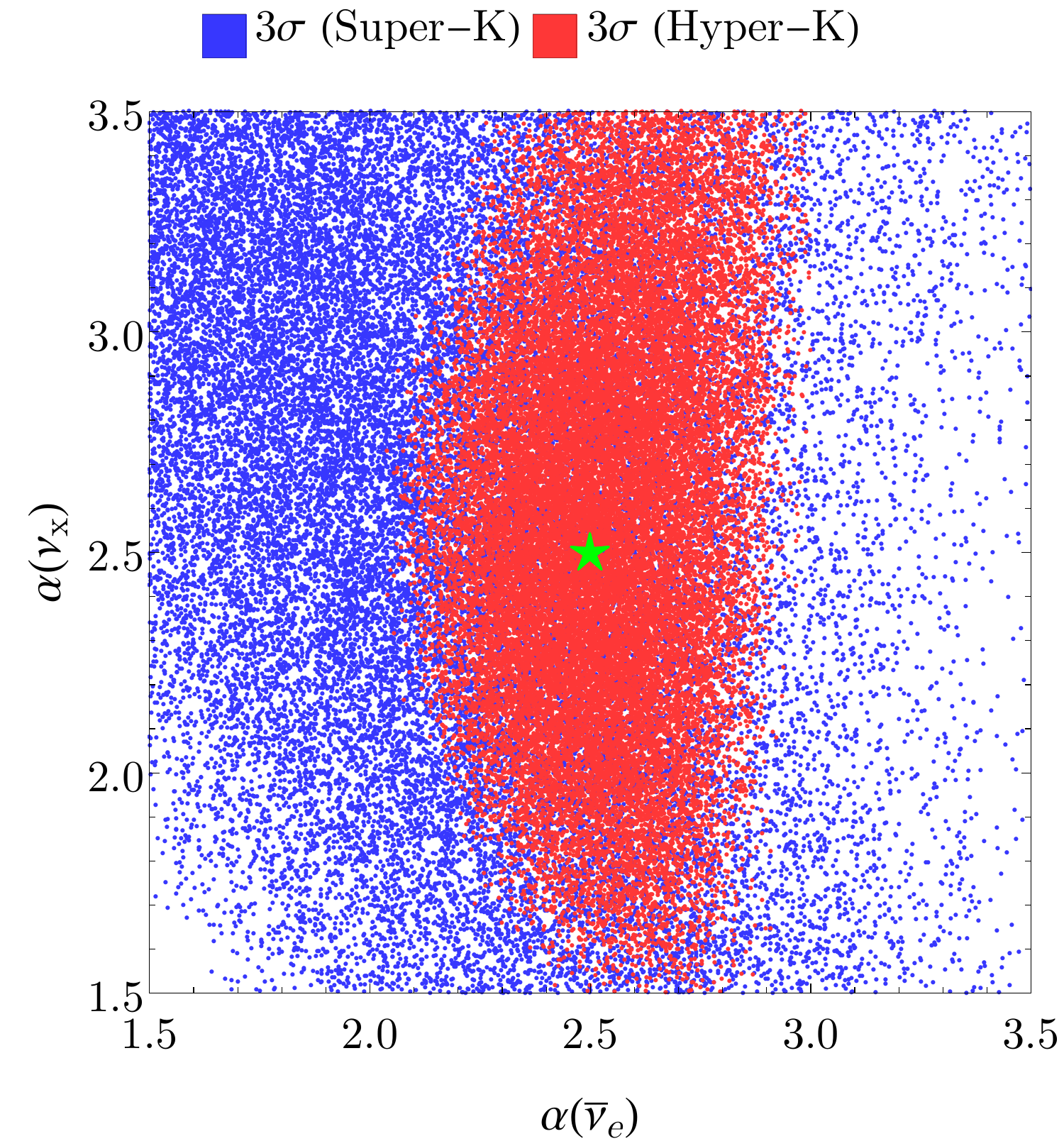}}
\caption{Projection of the extracted points onto
the $\alpha_{\APnue}$--$\alpha_{\Pnux}$
plane, in the IBD-only analysis 
\protect\subref{fig:scate3}
and combining all the detection channels 
\protect\subref{fig:scate4}. The blue dots refer to
Super-Kamiokande while the red ones to Hyper-Kamiokande.
Each set contains 30k points, accepted at $3\sigma$ CL.
The boundaries of the region are given by the prior
$[1.5,3.5]$ --- see section~\protect\ref{sec:1:2}.
The green star marks the true value.
}
\label{fig:pinc}
\end{figure}

As far as the pinching parameter is concerned,
figure~\ref{fig:pinc} shows the extracted points projected
onto the $\alpha_{\Pnue}$--$\alpha_{\Pnux}$ plane.
As one can see there, as well as in table~\ref{tab:risu},
the pinching distribution in Super-Kamiokande
is almost uniform inside the prior. Concerning Hyper-Kamiokande,
the combination of the three detection channels allows
to measure the pinching parameter of the
$\overline{\nu}_{\mathrm{e}}$ species
with a good accuracy ($\sim 7\%$). In fact this is expected: since
the total and mean energy are quite well determined, there
cannot be a large uncertainty on $\alpha_{\APnue}$.

Concerning $\alpha_{\Pnue}$, their values cover the 
prior range almost uniformly, and the $\sim23\%$
precision is simply due to its choice.
In fact, with a prior
uniform in the interval  $[1.5,3.5]$ we expect
a mean of 2.5 and a standard deviation of $1/\sqrt{3}$
that corresponds to a prior accuracy at the level of 23\%.
Such a result is close to the outcomes of most  available analyses,
and is what we find in ours (see table~\ref{tab:risu}).
Clearly, to pin down the $\alpha_{\Pnue}$ value more channels
sensitive to the electron neutrino spectrum should be added.

Finally, very similar conclusions on the average energies
and the pinching follow even assuming that an
exact equipartition of the neutrino emitted energies 
holds true. In fact, as one can see from table~\ref{tab:risu}, 
the reconstructed values of the average energies and pinching
parameters do not differ much without or with energy equipartition.

\section{Conclusions}
\label{sec:conclu}

In this manuscript we have focused on our ability
to reconstruct the neutrino spectra of a galactic supernova
in water Cherenkov detectors, i.e.\ Hyper-Kamiokande and 
Super\babelhyphen{-}Kamiokande.
Contrarily to most of the existing studies we have
not imposed any constraint to the flux parameters.
Also, we have combined here three detection channels, namely
inverse beta decay, elastic scattering on electrons and neutral
current scattering on oxygen.
Our 10 degrees of freedom likelihood analyses have shown that the
total and individual neutrino luminosities can be determined with
a few percent precision, except for the electron neutrinos.
In particular, the total emitted energy in Hyper-Kamiokande can
be reconstructed within a few percent ($\sim 6\%$), that becomes
less than $1\%$ in case the equipartition hypothesis holds true.
Note that information about this hypothesis is hardly extractable
from the data.

The average energies and pinching can also be identified at at few
percent level while the electron neutrino ones remain undetermined.
The best constrained by the analysis are the electron
antineutrinos: their mean energy and pinching can be
reconstructed up to $\sim2\%$  and $\sim7\%$ respectively in
Hyper-Kamiokande. Such results are mainly due to combining
information from inverse beta decay and elastic scattering.
They are also particularly remarkable because of
the absence of {\it a priori} choices of the flux parameters in the
likelihoods.  

The assumptions adopted in the present analysis can be extended in
several ways. For what concerns the neutrino fluxes, we have assumed
that they are given by quasi-thermal distributions, modified by the MSW
effect in normal ordering. Other cases can be studied, for instance,
those where the spectra undergo changes due to the neutrino
self-interactions. Future more precise inputs from the numerical
simulations for the fluxes and for their uncertainties are clearly
highly desirable. On the other hand, more detectors and/or detection
channels can be combined to those considered in the present work,
especially for the aim of reconstructing precisely the electron
neutrino fluxes, or for disposing of a clean sample of neutral current
events. The investigation of these issues will be the object of future
work.

The next (extra)galactic supernova is of great
interest for astrophysics and particle physics. 
In particular, although the uncertainties on the
neutrino fluxes are still large, this observation will bring crucial
information on the explosion mechanism, through the neutrino
light-curves, and on neutrino flavor conversion in the supernova,
with the reconstruction of the energy spectra.
Detailed investigations are needed to demonstrate that the neutrino
flux uncertainties will not prevent us from extracting important
information. The work presented here represents a supplementary
step in this direction.

\section*{Acknowledgments}
\addcontentsline{toc}{section}{Acknowledgments}

M.C.\ Volpe thanks K.\ Scholberg for providing useful information.
She also acknowledges financial support 
from ``Gravitation et physique fondamentale''
(GPHYS) of the {\it Observatoire de Paris}.
A.\ Gallo Rosso thanks the theory group of the
Astroparticle and Cosmology (APC) laboratory for
hospitality during the realization of the present work.

\end{document}